\documentclass[final,letterpaper,4pt,times,twocolumn]{elsarticle}

\journal{TexExchange}
    \journal{Journal of \LaTeX\ Templates}
    
    \usepackage{tgtermes}
    \usepackage{tikz}
    \usepackage{hyperref}
    \usepackage[letterpaper, portrait, margin=0.60in]{geometry}

    \usepackage{hhline}  
    \usepackage{float}
    \usepackage{subfig}
    \usepackage{multirow}
    \usepackage{graphicx,booktabs}
    \usepackage{xfrac}
    \usepackage{titlesec}
    \usepackage{wrapfig,booktabs}
    \usepackage{mwe}
    \usepackage{adjustbox}
    \usepackage{caption}
    \usepackage{makecell}
    \usepackage{rotating}
    
    \usepackage{amsmath,array, amssymb}
    \usepackage{xcolor}    
    \usepackage{longtable}
    \usepackage{ragged2e}

\hypersetup{colorlinks,breaklinks,
            urlcolor=Maroon,
            linkcolor=Maroon}
    \newcommand{\etal}{\textit{et al}.}
    \def\correction#1{%
    \abovedisplayshortskip=#1\baselineskip\relax\belowdisplayshortskip=#1\baselineskip\relax%
    \abovedisplayskip=#1\baselineskip\relax\belowdisplayskip=#1\baselineskip\relax}

    \fboxsep=\tabcolsep\relax
    \fboxrule=\arrayrulewidth\relax

    \newcolumntype{A}[2]{%
    >{\minipage{\dimexpr#1\linewidth-2\tabcolsep-#2\arrayrulewidth\relax}\vspace\tabcolsep}%
    c<{\vspace\tabcolsep\endminipage}}

\newcommand\bigzero{%
    \makebox[0pt]{\text{\normalsize Symmetric}}}
    
    \biboptions{numbers,sort&compress}

\begin{document}

    
    \begin{frontmatter}

    \title{\large Uncertainty Propagation in a Multiscale CALPHAD-Reinforced Elastochemical Phase-field Model}
    \author{Vahid Attari$^{a,*}$}
    \author{Pejman Honarmandi$^{a}$}
    \author{Thien Duong$^{a}$}
    \author{Daniel J Sauceda$^{a}$}
    \author{Douglas Allaire$^b$}
    \author{Raymundo Arroyave$^{a,b}$}
    \address{$^a$Materials Science and Engineering Department, Texas A\&M University, College Station, TX, USA 77840}
    \address{$^b$Mechanical Engineering Department, Texas A\&M University, College Station, TX, USA 77840}

    \cortext[mycorrespondingauthor]{Corresponding author email:attari.v@tamu.edu} 
    \small
    \begin{abstract} 
    ICME approaches provide decision support for materials design by establishing quantitative process-structure-property relations. Confidence in the decision support, however, must be achieved by establishing uncertainty bounds in ICME model chains. The quantification and propagation of uncertainty in computational materials science, however, remains a rather unexplored aspect of computational materials science approaches. Moreover, traditional uncertainty propagation frameworks tend to be limited in cases with computationally expensive simulations. A rather common and important model chain is that of CALPHAD-based thermodynamic models of phase stability coupled to phase field models for microstructure evolution. Propagation of uncertainty in these cases is challenging not only due to the sheer computational cost of the simulations but also because of the high dimensionality of the input space. In this work, we present a framework for the quantification and propagation of uncertainty in a CALPHAD-based elasto-chemical phase field model. We motivate our work by investigating the microstructure evolution in Mg$_2$(Si$_x$Sn$_{1-x}$) thermoelectric materials.  We first carry out a Markov Chain Monte Carlo-based inference of the CALPHAD model parameters for this pseudobinary system and then use advanced sampling schemes to propagate uncertainties across a high-dimensional simulation input space. Through high-throughput phase field simulations we generate 200,000 time series of synthetic microstructures and use machine learning approaches to understand the effects of propagated uncertainties on the microstructure landscape of the system under study. The microstructure dataset has been curated in the Open Phase-field Microstructure Database (OPMD),  available at \href{http://microstructures.net}{http://microstructures.net}.
    \end{abstract}


    \end{frontmatter}
    
 
    
    \section{Introduction}\label{1}
    
    \fontdimen2\font=0.5ex
    \justify

Uncertainty Quantification (UQ) has a long and successful history of application to very diverse areas such as climate change \cite{qian2016uncertainty}, structural engineering \cite{gupta2017uncertainty}, aerospace engineering and design~\cite{ghoreishi2017adaptive}, and medicine \cite{begoli2019need}, to name a few. In the field of materials science, however, notions of UQ remain relatively unexplored even though it is of critical importance as the field progresses towards more quantitative/predictive approaches to materials development. Indeed, uncertainty quantification (UQ) and its propagation (UP) across model/simulation chains are key elements of decision-based~\cite{panchal2013key,mcdowell2016materials,arroyave2018interdisciplinary,arroyave2019systems} materials design in the framework of Integrated Computational Materials Engineering (ICME)~\cite{allison2011integrated}. The latter prescribes the integration of databases, multi-scale modeling and experiments with the aim to reduce the time and effort of the materials development cycle~\cite{liu2018vision}. Given the complexity and computational cost of most materials simulation frameworks, it is necessary to have a systematic approach to quantify uncertainties in the parameters/variables in any system of interest and to propagate these uncertainties to the respective responses of individual or multi-scale systems. 

Despite the importance of UQ/UP in multi-scale modeling \cite{weinan2011principles,sanghvi2019uncertainty}, there are very few works in the literature dealing with uncertainty quantification (UQ) and/or propagation (UP) across multi-scale models in the field of materials science and engineering. Liu et al. \cite{liu2009complexity}, for example, focused on the probabilistic prediction of the effective properties in heterogeneous composite materials and their performance. In that work, UQ of the parameters and UP across the multi-scale constitutive models (i.e. UP from structure to property to performance) were performed through a Bayesian stochastic method and a stochastic projection technique, respectively. Some works for UP across the multi-scale modeling for the probabilistic predictions of plastic flow behavior in poly-crystalline materials have been described thoroughly in \cite{kouchmeshky2010microstructure,koslowski2011uncertainty,salehghaffari2012new}. Recently, Honarmandi \etal~\cite{honarmandi2019uncertainty_NITI} demonstrated the use of UQ approaches to the parameterization of thermodynamically rigorous models for the response of NiTi-based shape memory alloys, followed by uncertainty propagation over the model parameter space. 

In computational materials science, thermodynamic assessments using the CALculation of PHAse Diagrams (CALPHAD) method~\cite{saunders1998calphad} constitutes the basis for a broad range of approaches to materials simulations, including microstructure evolution through phase field modeling~\cite{steinbach2007calphad}. Given the foundational nature of CALPHAD-based descriptions of phases' free energies in any attempt to predict processing-(micro)structure relationships, UQ/UP in CALPHAD~\cite{stan_2003} play a very important role, although sparse examples in the literature address this. Honarmandi \etal~\cite{honarmandi2019bayesian} used a Bayesian framework to quantitate and propagate uncertainty in the context of CALculation of PHAse Diagrams (CALPHAD) thermodynamic assessments and showed how information fusion approaches~\cite{thomison2017model} can be used to fuse propagated uncertainties from different competing models. Other groups have also demonstrated different frameworks for the quantification and propagation of uncertainty in CALPHAD models~\cite{honarmandi2019uncertainty,otis2017high,bocklund2019espei}. 

Similarly to the case of CALPHAD-based thermodynamic assessments, the application of UQ/UP frameworks to phase-field modeling remains relatively unexplored~\cite{kim1999phase,steinbach2012phase,attari2018interfacial,attari2018exploration,karayagiz2019finite,duong2019probing}. Koslowski \etal~\cite{koslowski2011uncertainty} characterized how  uncertainties propagate across spatial and temporal scales in a physics-based model of nanocrystalline plasticity of fcc metals, combining molecular dynamics (MD) with phase field dislocation dynamics (PFDD) simulations. Wang \etal~\cite{wang2014asymptotic} carried out an asymptotic and uncertainty analysis of void formation during irradiation. Leon \etal~\cite{leon2017identifiability} used subset selection and active subspace techniques to identify dominant parameters in a continuum phase-field poly-domain model for ferroelectric materials. These earlier approaches focused on UQ/UP over a single modeling framework, but B{\"o}ttger~\cite{bottger2018icme} recently demonstrated the propagation of uncertainty across an entire ICME-based model chain. 

Across different fields~\cite{lee2009comparative}, UP is practically implemented through many different approaches, including Monte-Carlo-, local expansion-,functional expansion-, and numerical integration-based methods. By far, the most basic and common approach to propagating uncertainty through computational models is via Monte Carlo simulation~\cite{honarmandi2018parametric}. For expensive computational models, however, the use of sample-based approaches are often computationally prohibitive. While one can use surrogate models to efficiently sample the input/output relationships~\cite{tapia2017bayesian}, numerical efficiency often comes at the expense of fidelity. Moreover, such approaches tend to fail in cases in which the model output changes qualitatively (not only quantitatively) in different regions of the input space. Phase-field simulations belong to the class of computational problems where these conventional approaches to UP tend to be ineffective. The challenges associated with the computational expense and the non-regular nature of the output of phase field simulations are augmented by the large dimensionality of the input space. Finally, the complexity of the output of phase-field simulations makes the analysis of UP frameworks challenging.

The major objective of the current work is to show the propagation of statistically quantified uncertainties of thermodynamic parameters to the Gibbs free energy of phases, and equilibrium phase diagram, in combination with the uncertainties of microelastic and kinetic parameters. Consequently, uncertainties in the Gibbs free energies upto microstructural characteristics 
are studied by using high throughput analyses across a chain of models that includes a CALPHAD, a microelasticity, and a phase field model, consecutively. The uncertainty of thermodynamic parameters in the CALPHAD model is quantified through a Markov Chain Monte Carlo (MCMC) sampling technique in the context of Bayesian statistics, while the uncertainty of microelastic and kinetic parameters is determined through prior knowledge, calculations and/or estimations.
    
\begin{figure*}[!ht]
        \centering
        \includegraphics[width=\textwidth]{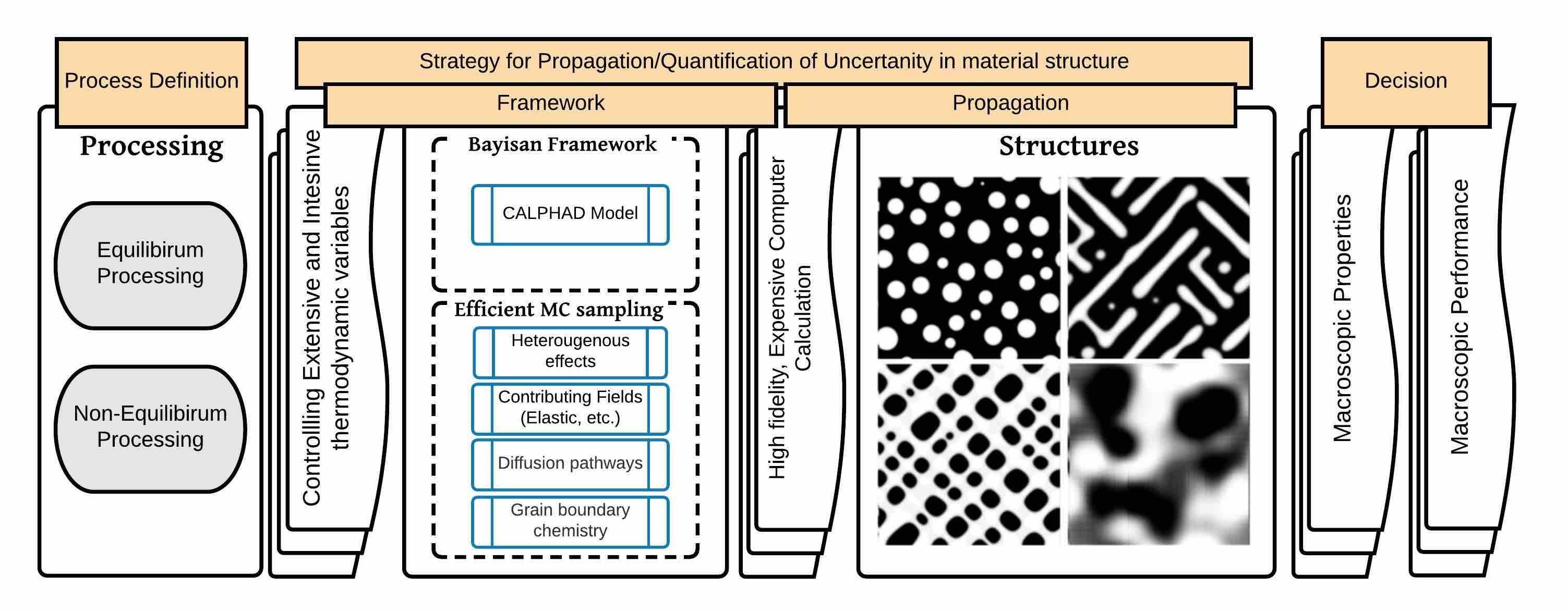}
        \caption{The process-structure-property uncertainty propagation framework deployed in composition, strain and temperature space based on the process definition and natural uncertainties in input parameters.}
        \label{fig:Alloy_Thermo}
    \end{figure*}

Figure \ref{fig:Alloy_Thermo} demonstrates the proposed strategy and steps toward developing a framework for propagating the uncertainty across CALPHAD-phase field model chains. The steps are based on determining 1) type of processing conditions, 2) interacting physics (sub-models), 3) Quantities of Interest (QoIs), and 4) microscopic/macroscopic properties of interest. We propose to demonstrate the framework in the investigation of the microstructure evolution of nanostructured Mg$_2$(Si$_x$Sn$_{1-x}$) thermoelectric (TE) materials\cite{wang2009improved,yi2018strain}. This pseudo-binary system is characterized by a miscibility gap between two isomorphous cubic phases~\cite{yi2018strain}. The tendency of this system to phase-separate or homogenize is taken into account by employing a fully parameterized elasto-chemical phase-field model that accounts for the effect of process conditions on the resulting microstructure configurations. 

Since the performance of multi-phase thermoelectric materials is greatly determined by their (multi-scale) microstructure, it is expected that changes in processing schemes may have considerable impact on performance. In fact, for the case of Mg$_2$(Si$_x$Sn$_{1-x}$)-based thermoelectrics, it has been already shown that equilibrium and non-equilibrium processing can lead to dramatic changes to the TE figure of merit, ZT~\cite{yi2018strain,aizawa2006mechanically,noda1992preparation}. Understanding of this TE system is further hindered by the considerable uncertainty in the location of the miscibility gap, with different experimental phase boundary estimations disagreeing by several tens of atomic percent~\cite{yi2018strain}.

In this work, we attempt to propagate uncertainty through phase field simulations, accounting for the uncertainty not only in the parameters directly associated to the phase field model, but also arising from uncertainty in the CALPHAD parameterization. This paper is structured as follows: in section \ref{sec:TE_materials}, we motivate the present work through the application of ICME-based frameworks on the design of nanostructured TE materials. We note that the framework put forward is generalizable to a wide range of materials problems. The details of the models (CALPHAD and phase-field) and uncertainty propagation strategy is provided in section \ref{sec:Methods}. Section \ref{sec:results_Discussion} discussed the prior and the resulting uncertainty in the phase diagram, the subsequent uncertainty in the microstructures under elasto-chemical simulations, and the methods for data interpretation/classification. Furthermore, we present a summary on our findings and draw our conclusions in section \ref{sec:conclusions}.

\section{Nanostructured Thermoelectric Materials}  
\label{sec:TE_materials}
As mentioned above, the motivating example is the propagation of uncertainty in microstructure evolution simulations in nanostructured composite thermoelectrics (TE)~\cite{heinz2014applying,balout2015thermoelectric,gorsse2016multi,gorsse2009microstructure} via direct coupling of CALPHAD thermodynamic assessments~\cite{wang2018design} with multi-physics phase field models (PFM)~\cite{attari2018interfacial,yi2018strain} that account for both chemical and elastic driving forces for structure formation. The example is motivated by recent work by some of the present authors~\cite{yi2018strain} on the dramatic effect that processing has on the microstructure (and TE performance) in Mg$_2$Si-Mg$_2$Sn alloys, but \textbf{has a much broader applicability} as CALPHAD/PFM-based microstructure simulations are pervasive in ICME-based frameworks for microstructure-sensitive materials design~\cite{furrer2011application,schmitz2009toward,yabansu2017extraction,wang2014integrated,schmitz2015microstructure}, and properly accounting for uncertainty is necessary to make design choices with proper confidence bounds.

Current interest in the thermoelectric (TE) effect  originates from the ever increasing demand for energy and the associated detrimental effects on global climate. Current TE materials, unfortunately, do not have the efficiency---described by the figure of merit $z^T=\alpha^2sT/\alpha$, where $\alpha$ is the Seebeck coefficient, $s$ is the electrical conductivity, T is the absolute temperature and $\kappa$ is the thermal conductivity---that would turn TE-based devices into competitive power-generators~\cite{heinz2014applying,snyder2011complex}. An ideal TE material would have a large Seebeck coefficient, while being electrically conductive and thermally insulating~\cite{snyder2011complex,heinz2014applying}. These properties, however, are coupled and their individual tuning is thus challenging. Over the past decade, a sophisticated arsenal of strategies for the rational design of TE materials has emerged~\cite{yang2013rational,tan2016rationally,gorai2016te}, including the exploitation of spontaneous self-assembly or non-equilibrium processing of nanostructures to enhance  phonon-scattering~\cite{heinz2014applying,yi2018strain}.

Among the hundreds of TE systems investigated to date, environmentally-benign Mg$_2$(Si$_x$Sn$_{1-x}$) alloys~\cite{tazebay2016thermal,liu2012convergence} have attracted considerable attention due to their relatively high figure of merit ($zT>1$)~\cite{zhang2008high}, comparable with the intermediate temperature TE materials such as PbTe and filled skutterudites~\cite{zaitsev2006highly,liu2012convergence,nolas2007transport,yi2018strain}. The Mg$_2$(Si$_x$Sn$_{1-x}$) pseudo-binary system exhibits a miscibility gap~\cite{vives2014combinatorial,kozlov2011phase,viennois2012phase,nikitin1961thermoelectric} and this has been exploited to realize nanostructures with optimal TE performance~\cite{tazebay2016thermal,polymeris2015nanostructure,zhang2008high}. Experimental determination of ZT even in a single alloy exhibits considerable variance, perhaps due to changes in the way these materials are synthesized and processed. Recently, the present authors and collaborators investigated the effect of non-equilibrium processing on the  microstructure evolution (and transport properties) in the Mg$_2$Si$_{0.7}$Sn$_{0.3}$ system and found that instead of phase-separating, the system tended to form a solid-solution with superior TE performance, contrary to expectations and prior works~\cite{tazebay2016thermal,polymeris2015nanostructure}. This was ascribed to (elastic) coherency effects and was \emph{\textbf{verified} via quantitative multi-physics phase field simulations}~\cite{yi2018strain}.

 
 These results are interesting as they exemplify the influence of processing on the microstructural evolution in TE materials~\cite{heinz2014applying} and the corresponding change in performance. Most importantly, this constitutes one of the very few examples---to the best of our knowledge---in which phase-field modeling (PFM), in combination with CALPHAD free energies has been used to investigate the microstructure evolution of TE materials. Further investigation of the PFM developed to study the Mg$_2$Si$_{0.7}$Sn$_{0.3}$ showed that rather small changes in the strength of the elastic couplings---mediated via lattice parameter differences between Si- and Sn-rich domains---resulted in qualitatively different microstructures, which in turn could be expected  to exhibit different phonon transport behavior. 

The ability to quantitatively understand~\cite{heinz2014applying} and control the different materials and processing parameters related to microstructural morphology, topology, size and spacing in composite TE materials has already been demonstrated~\cite{rowe1981phonon,medlin2009interfaces,ikeda2007self,pei2011combination,gorsse2009microstructure}. Many of these approaches have been inspired by metallurgy and thus the \textbf{time is ripe} to translate much of what has been learned on ICME-enabling microstructure-sensitive (structural) alloy design to the problem of designing (self-assembled) TE microstructures for optimal performance. While the modeling framework via PFM has shown to result in (semi-)quantitative predictions that compare well with experiments~\cite{yi2018strain}, a robust ICME research program on microstructure design of TE materials requires reliable and efficient UQ/\textbf{UP} frameworks.
    
\begin{figure}[!ht]
        \centering
        \includegraphics[width=0.4\textwidth]{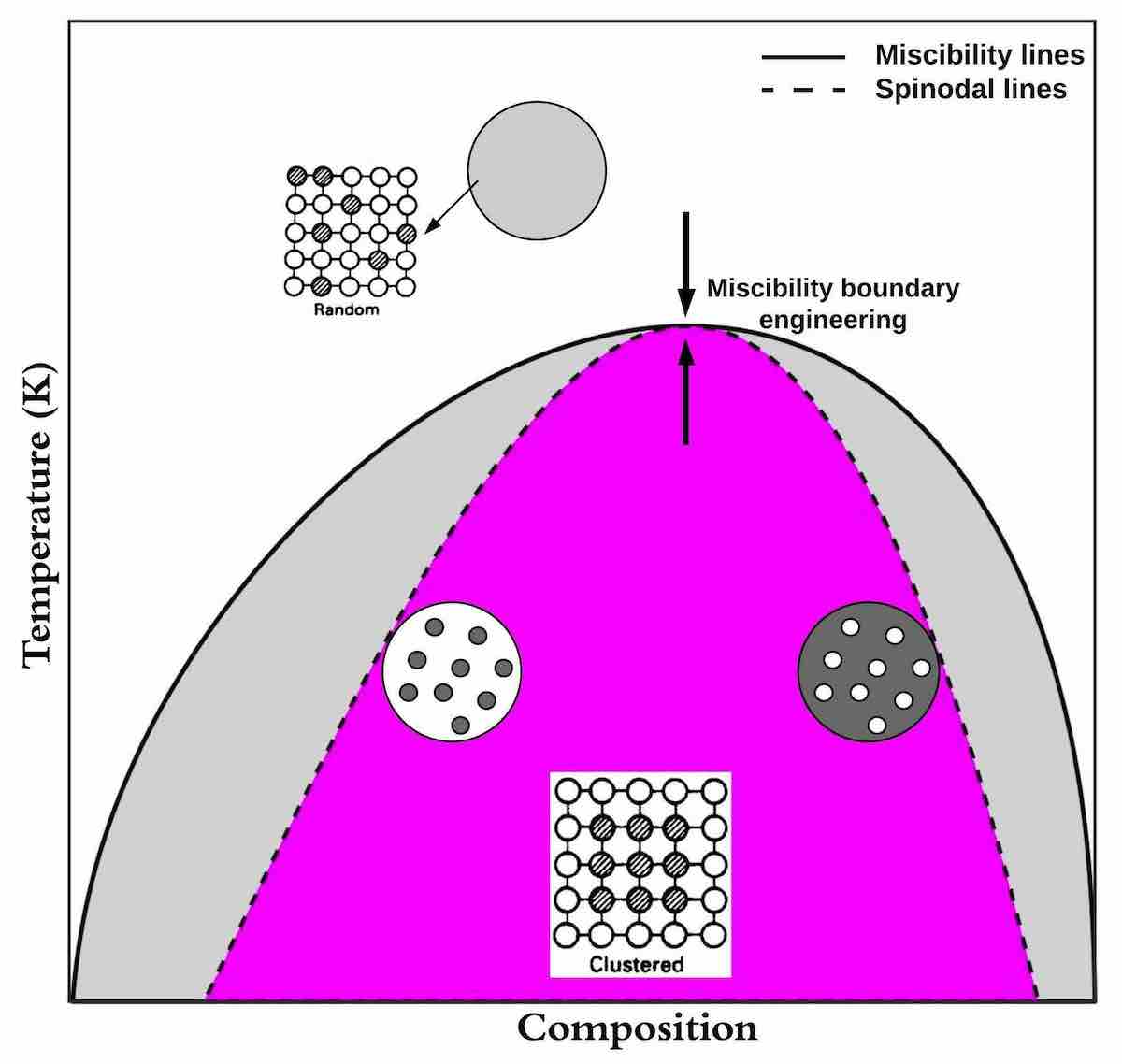}
        \caption{A general strategy to enhance thermoelectric performance by engineering better phonon scattering via nanoscale precipitates or solid solution.}
        \label{fig:strategy_1}
    \end{figure}

Figure \ref{fig:strategy_1} illustrates an schematic phase diagram in which the material shows an inherent instability in certain regions of the composition space. This material is uniform at high temperatures and upon reducing the temperature decomposes into distinct phases. Similarly, Mg$_2$X\{Sn,Si\} system has a miscibility where the uncertainties in the Sn-rich and Si-rich boundaries (Refer to Fig. 3 of \cite{yi2018strain}) has been the center of discussions in several articles \cite{wang2009improved,yi2018strain,viennois2012phase,vivès2014combinatorial,kozlov2011phase}. As a consequence, these uncertainties impact the chemical free energy and (the predicted) microstructure of this system. 

The extent of instabilities, the uncertainty in the boundaries of the unstable region, and extensive possibilities by means of using different synthesis conditions all suggest that analysis and quantification of uncertainty is necessary in this material to establish the correlation between the thermodynamics and microstructural phenomena. These analyses are essential to estimate how the variance in processing and (epistemic or aleatoric) uncertainties in material parameters can affect the microstructure and ultimately the TE response of this material. This also can help to better realize the reliable thermodynamic conditions for enhancement of the properties of TE materials \cite{gorai2017computationally}.

\section{Models and Methodologies} 
\label{sec:Methods}
    
Here, we first define the thermodynamic state variables relevant to the elasto-chemical phase field model formulation and then quantify the uncertainties in the given material variables and propagate these uncertainties with the aim of quantifying the variations in the micro-structure of the material that ultimately alters the predicted macroscopic response.
    
\subsection{Thermodynamic State Variables}
\label{sec:thermodynamic_state_variables}

For an isothermal and isobaric state, the total free energy functional ($F^{tot}$) for an undeformed material configuration with confined boundaries can be constructed from the sum of all contributing fields over it. We restrict $F^{tot}$ to chemical and elastic contributing effects here. Consequently, the energy is a function of composition ($c$), temperature ($T$), strain ($\varepsilon$), gradients of composition ($\nabla c$), and other fields, if they were present. We write the total free energy functional, $F^{tot}$ as the sum of contributing fields:
    
    \begin{equation}\label{F_tot}
        F^{tot}(c,T,\nabla c,\varepsilon) = f^{bulk} + f^{interfacial} + f^{elas}
    \end{equation}
    
    \noindent
    where the bulk free energy density, $f^{bulk}$, the interfacial free energy density, $f^{interfacial}$, and elastic strain energy density, $f^{elas}$ are:
    
    \begin{equation}\label{F_chem}
        f^{bulk}(c,T) = \int_\Omega \left( f^{0}(c,T)  \right ) d\Omega
    \end{equation} 
    \vspace{-0.5cm}
    \begin{equation}\label{F_int}
        f^{interfacial}(\nabla c) = \int_\Omega \kappa(\nabla c)^2  d\Omega
    \end{equation}    
    \vspace{-0.5cm}
    \begin{equation}\label{F_elas}
        f^{elas}(\varepsilon) = \int_\Omega \sigma_{ij} \varepsilon^{el}_{ij} d\Omega
    \end{equation}
    
    \noindent
where $f^{0}(c,T)$ is the free energy of a unit volume of homogeneous material, $\kappa$ is the gradient energy coefficient, $\varepsilon^{el}_{ij}$ and $\sigma_{ij}$ are the local elastic strain and stress in the material, respectively. The chemical free energy is composed of the interfacial and bulk energy contributions, and it determines the compositions and volume fractions of the equilibrium phases. The strain energy affects the equilibrium compositions and volume fractions of the coexisting phases, but also determines the shapes and configurations of the phase domains.

The bulk free energy of the Mg$_2$(Si$_x$Sn$_{1-x}$) pseudo-binary system is described through the sub-regular solution model as:

    \footnotesize
    \begin{equation}\label{eq 1}
    f^{bulk}(c_{\alpha},T) = \sum_{{\alpha}} c_{\alpha} G_{\alpha}^0 + RT\sum_{\alpha}{c_{\alpha} ln(c_{\alpha})} + \sum_{{\alpha}} \sum_{\beta>{\alpha}}c_{\alpha} c_{\beta} \sum_{\nu} L_{\alpha\beta}^{\nu} (c_{\alpha}-c_{\beta})^{\nu}
    \end{equation}
    \normalsize
    
    
\noindent
where $c_{\alpha}$ is the mole fraction of the constituent $\alpha$ (either Mg$_2$Sn or Mg$_2$Si), $G^{^0}_{\alpha}$ is the reference energy of the constituent, $T$ is temperature (in K), $R$ is the gas constant, and $L_{\alpha\beta}^{\nu}$ is given as:

    \begin{equation}
    \label{eq 2}
    L_{\alpha\beta}^{\nu} = {^\nu}a_{\alpha\beta} + {^\nu}b_{\alpha\beta}T
    \end{equation}
    
    \noindent

where ${^\nu}a_{\alpha\beta}$ and ${^\nu}b_{\alpha\beta}$ are model parameters, which describe the interactions between the constituents beyond those of ideal mixing. These parameters can be calibrated against the available data deterministically or probabilistically, as discussed later. In order to build the phase diagram, the total Gibbs free energy of the system is minimized at different temperatures for volume fraction and composition of each existing phase.
    
\subsection{Material Modeling Strategy: CALPHAD Reinforced Phase-field Method}\label{3.2}
    
From linear kinetic theory, the local mass flux in the presence of a gradient in composition, the diffusion flux, $\vec{J}$ (in units of $mol~m^{-2}~s^{-1}$) is given by:
    
    \begin{equation}
        \vec{J} = -\vec{M} \nabla \mu^{tot}
    \end{equation}    
    
    \noindent
    where $\vec{M}$ is the interface mobility assumed to be constant due to the isotropic nature of the crystal structures of the two phases, and $\mu^{tot}=\frac{\delta F^{tot}}{\delta c}$ is the total potential for the kinetic transition. We postulate the following form of the Cahn-Hilliard (C-H) kinetic equation along with other micro-elasticity equations to study the evolution of the Mg$_2$(Si$_x$Sn$_{1-x}$) microstructure in space:
    
    \begin{equation} \label{eqn:C_Heqn}
        \frac{\partial c}{\partial t} = \nabla \vec{M} \nabla \left ( \frac{\delta F^{tot}}{\delta c} \right )
    \end{equation} 
    \vspace{-0.5cm}
    \begin{equation} \label{eqn:mech_eq}
        \frac{\partial \sigma_{ij}}{\partial r_j} = 0 
    \end{equation}     
    \vspace{-0.5cm}
    \begin{equation} \label{eqn:el_strain}
        \varepsilon_{ij} = \frac{1}{2} \left ( \frac{\partial u_i}{\partial r_j} - \frac{\partial u_j}{\partial r_i} \right )  
    \end{equation}     
    \vspace{-0.5cm}
    \begin{equation} \label{eqn:hookslaw}
        \sigma_{ij} = C_{ijkl}\varepsilon^{el}_{kl} 
    \end{equation}
    \vspace{-0.5cm}
    \begin{equation} \label{eqn:eps_tot}
        \varepsilon^{el}_{kl} = \varepsilon^{tot}_{kl} - \varepsilon^{0}_{kl} 
    \end{equation}

Equations~\ref{eqn:C_Heqn}-\ref{eqn:eps_tot} are the C-H equation, mechanical equilibrium condition, kinematics, Hooke's microscopic constitutive law for linear elasticity, and strain relationship, respectively. The dilatational eigenstrain term is given by $\varepsilon^{0}_{kl} = \varepsilon^{T}\delta_{kl}h(c)$, and it is the consequence of lattice strain between the phases. $\delta$ is the Kronecker-delta function and $h(c)$ is an interpolation function. $C_{ijkl}$ is the composition-dependent fourth order elastic modulus tensor. It is convenient to describe $C_{ijkl}$ using the following expression:
    
    \begin{equation}
        C_{ijkl}(c) = C^{eff}_{ijkl} - g(c)\Delta C_{ijkl}
    \end{equation}
    
    \noindent    
    where $\Delta C_{ijkl} = C^{\alpha}_{ijkl} - C^{\beta}_{ijkl}$ is the difference between the elastic moduli tensor of $\alpha$ and $\beta$ phases. $C_{ijkl}$ is a $6\times6$ symmetric tensor which can be denoted by the Voigt notation for any linearly elastic medium in the form of:
    
    \begin{equation}
        \scriptsize
        \text{C}_{ijkl} \rightarrow \text{C}_{mn} = \left[ 
        {\begin{array}{cccccc}
            \text{C}_{11} & \text{C}_{12} & \text{C}_{13} & \text{C}_{14} & \text{C}_{15} & \text{C}_{16} \\
                          & \text{C}_{22} & \text{C}_{23} & \text{C}_{24} & \text{C}_{25} & \text{C}_{26} \\
                          &               & \text{C}_{33} & \text{C}_{34} & \text{C}_{35} & \text{C}_{36} \\
                          &               &               & \text{C}_{44} & \text{C}_{45} & \text{C}_{46} \\
                          & \bigzero      &               &               & \text{C}_{55} & \text{C}_{56}\\
                          &               &               &               &               & \text{C}_{66}\\
        \end{array}}
        \right]
    \end{equation}

Similar to $h(c)$, $g(c)$ is a scalar-valued interpolation function such that $h(c_{\alpha})=g(c_{\alpha})=0$ and $h(c_{\beta}) = g(c_{\beta})=1$. The phase-field microstructural evolution problem (eqn. \ref{eqn:C_Heqn}) is solved by utilizing a semi-implicit Fourier spectral approach~\cite{chen1998applications}. The microelasticity problem (eqns. \ref{eqn:mech_eq}-\ref{eqn:eps_tot}) is solved by the FFT-based iterative solver described in \cite{attari2018exploration}. \textcolor{black}{In this approach, the following conditions are considered: 1) stress-free transformation strains (SFTS) for each phase and inhomogeneous elastic constants in the domain, 2) $\Delta C_{ijkl}=0$ in the first iteration, 3) strain-control based on stress-control, 4) convergence of the problem when the $L^2$ norm of $||\boldsymbol{u}^{n+1}-\boldsymbol{u}^{n}||$ is less than $10^{-8}$, and 5) Periodic boundary conditions in all sides of the domain.}

The parameter space for CALPHAD, micro-elasticity, and phase-field models is listed in Table \ref{table:1st_order_elastic_cts}.
     
    
    \begin{table*}[!ht]                                       
        \centering 
        \scriptsize
        \caption{The quantification of uncertainty in the CALPHAD model and 1$^{st}$ order statistics on the outputs of inputs of the phase-field model. The outputs of the CALPHAD model were fed to the phase-field model. The fitted truncated PDFs which correspond to the plausible optimal values and uncertainty of the model parameters. "ss" and "liq" denote Mg$_2$(Si$_x$Sn$_{1-x}$) and liquid phases, respectively. ** SFTS stands as Stress-Free Transformation Strain.}                            
        \begin{tabular}{m{0.5cm}m{1.25cm}m{3.05cm}m{2cm}m{1cm}m{1.5cm}m{1.75cm}m{1.50cm}m{1.50cm} @{}m{0cm}@{}}  
        \toprule                                                
        Model & Sub-model & Parameter & PDF form & Unit & $\mu$ & Dispersion ($\frac{\sigma^2}{\mu}$) & Lower Bound & Upper Bound \\    
        \midrule 
        

        & & $^{0}a_{ss}$  & {\includegraphics[scale=0.055]{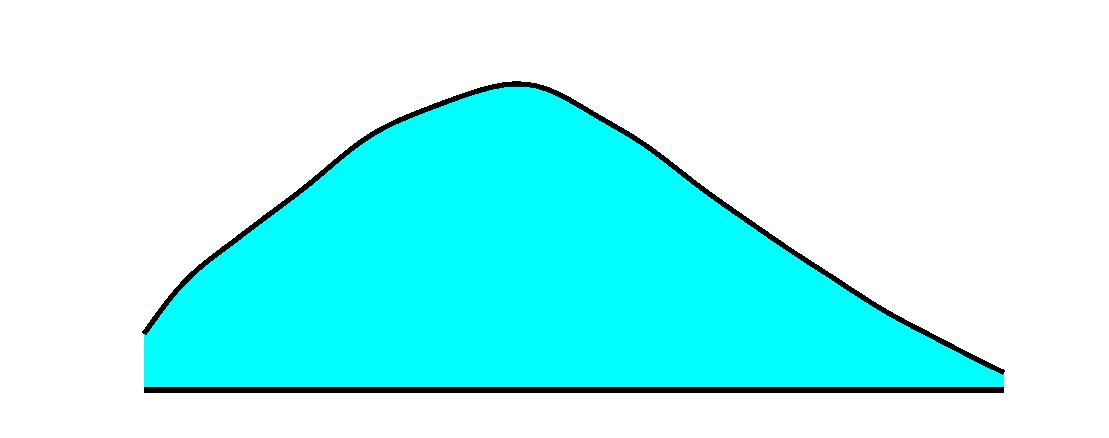}} & $J.mol^{-1}$ & 12840.4 & 729.0 & 6824.89 & 20474.69 \\   
        & \parbox[c]{2mm}{\multirow{1}{*}{\rotatebox[origin=c]{90}{\normalsize{CALPHAD Model Parameters}}}} & $^{0}b_{ss}$  & {\includegraphics[scale=0.055]{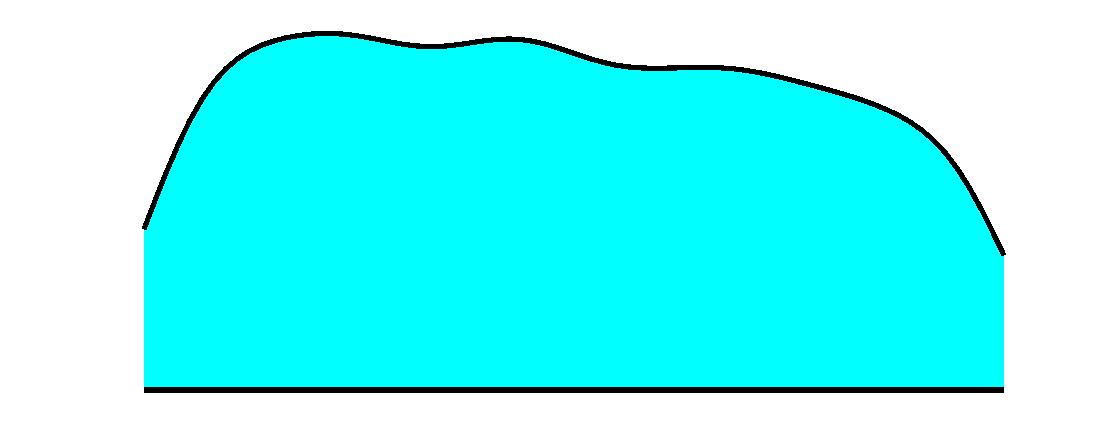}}  & $J.mol^{-1}$ & 7.20 & 0.59 & 3.67 & 11.02 \\   
        & & $^{1}a_{ss}$  & {\includegraphics[scale=0.055]{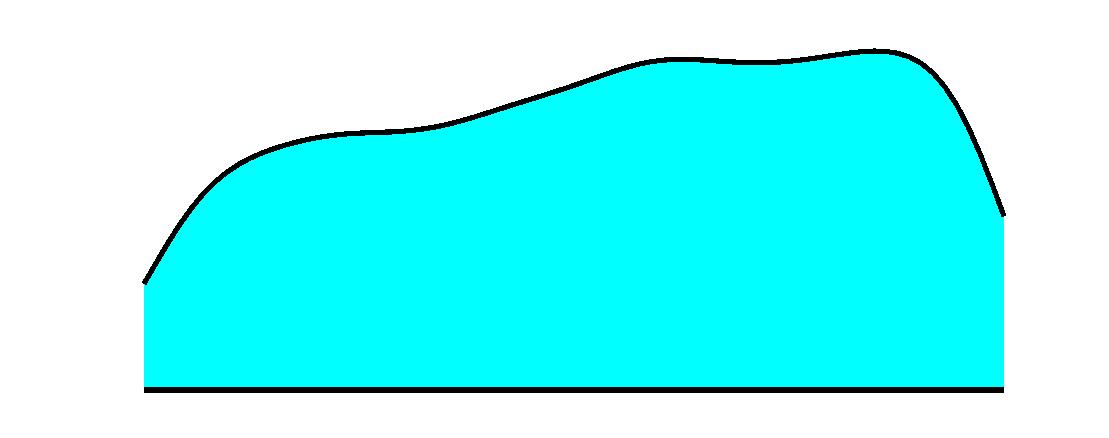}}  & $J.mol^{-1}$ & -3324.3 & -288.3 & -5208.34 & -1736.11 \\   
        & & $^{0}a_{liq}$ & {\includegraphics[scale=0.055]{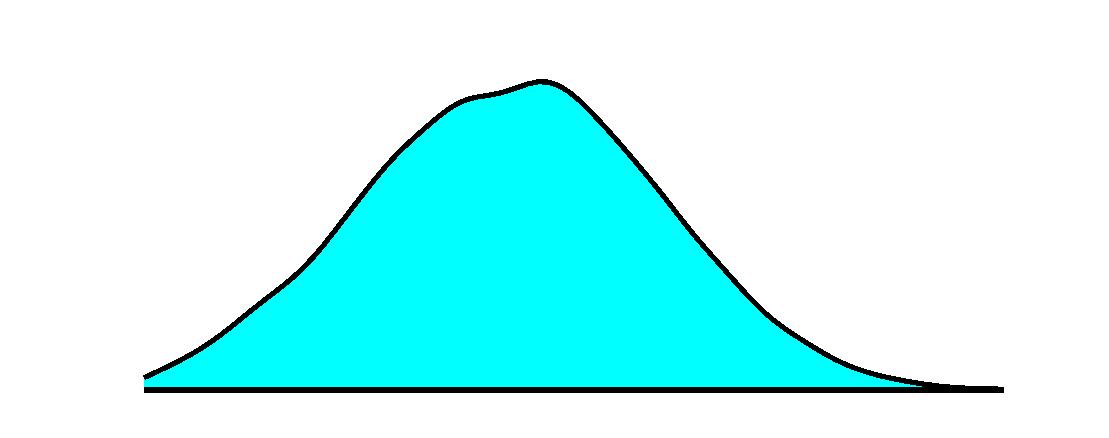}}  & $J.mol^{-1}$ & 80635.7 & 26688.6 & 43550.19 & 130650.57 \\
        \parbox[c]{2mm}{\multirow{1}{*}{\rotatebox[origin=c]{90}{\large{Phase-field Model Parameters}}}} & & $^{0}b_{liq}$ & {\includegraphics[scale=0.055]{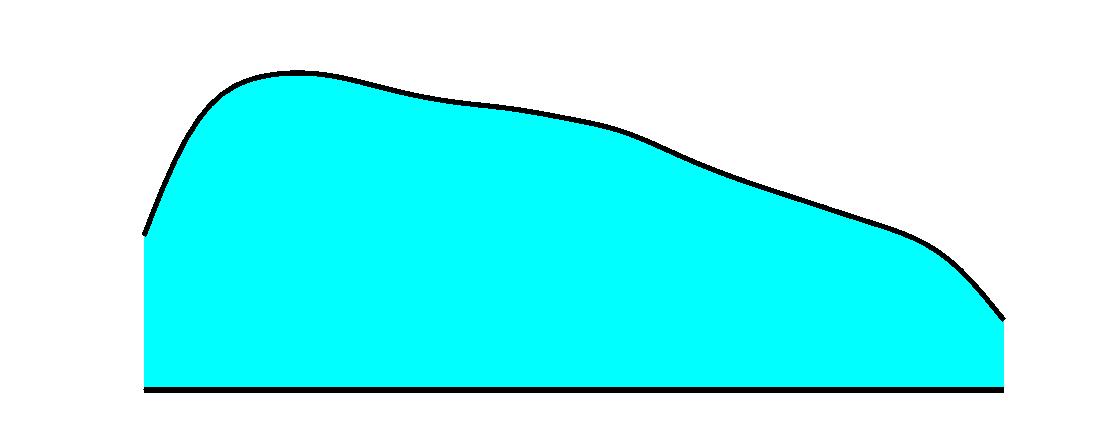}}  & $J.mol^{-1}$ & -61.25 & 3.98 & -86.03 & -28.68 \\    
        & & $^{1}a_{liq}$ & {\includegraphics[scale=0.055]{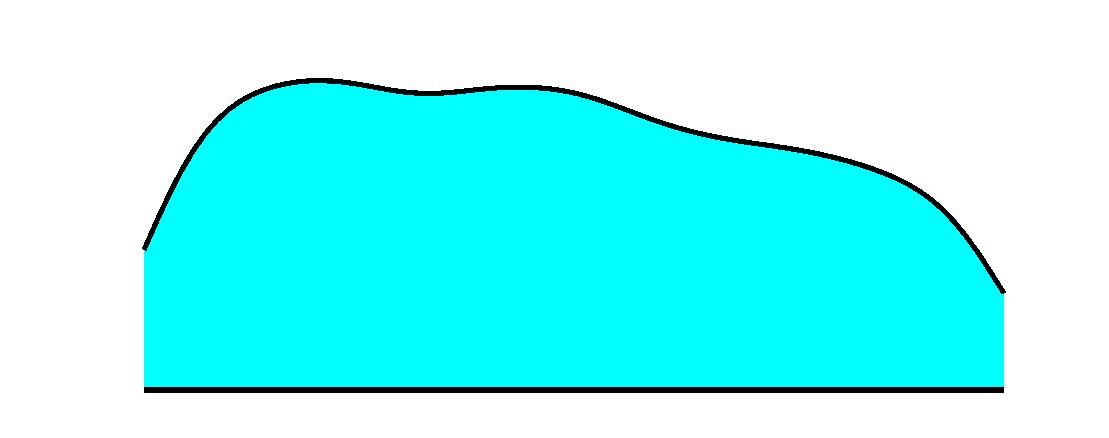}}  & $J.mol^{-1}$ & 6400.5 & 529.01 & 3314.80 & 9944.41 \\ \cmidrule(lr{1em}){3-3}

        & & Alloy composition ($c$) & {\includegraphics[scale=0.055]{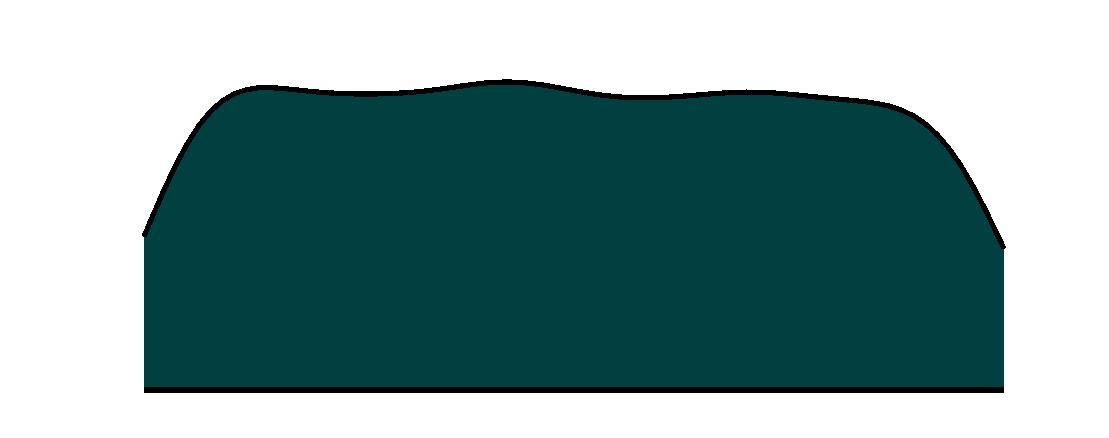}} & $mol$ & 0.40 & 0.0083 & 0.3 & 0.5                               \\ \cmidrule(lr{1em}){3-3}
        
        & \parbox[c]{4mm}{\multirow{1}{*}{\rotatebox[origin=c]{90}{\normalsize{Kinetic Parameters}}}} & Interface mobility $(M)$ & {\includegraphics[scale=0.055]{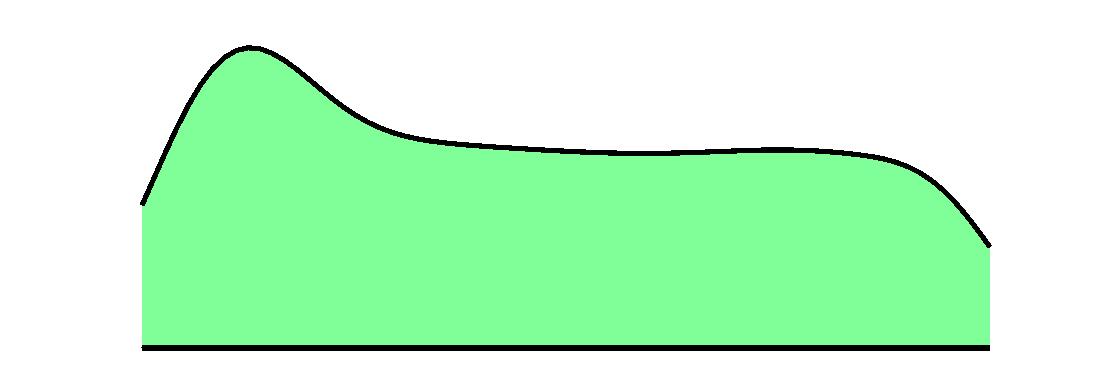}}  & $m^2s^{-1}.J^{-1}$ & $5.62\times10^{-19}$ & $4.64\times10^{-25}$ & $10^{-18}/(RT)$ & $10^{-20}/(RT)$ \\    
        & & Gradient energy Coefficient $(\kappa)$ & {\includegraphics[scale=0.055]{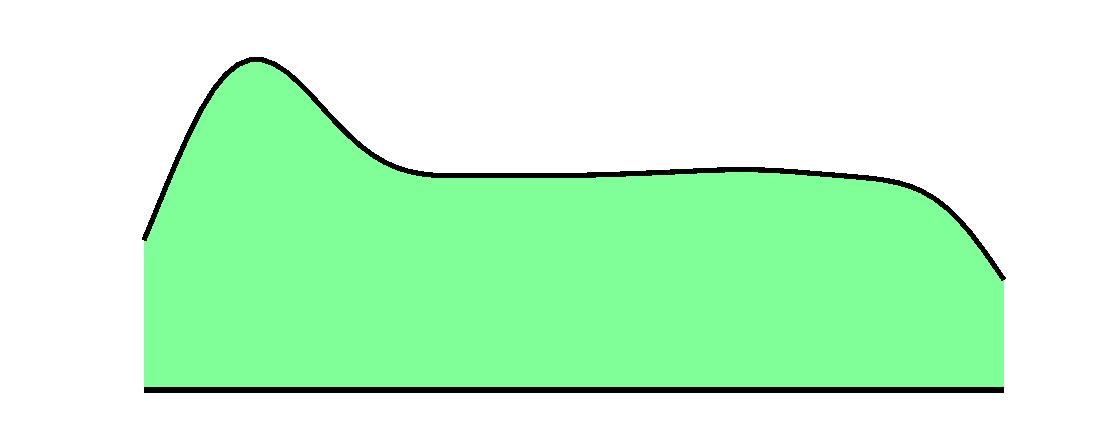}} & $J.m^{-2}$ & $1.16\times10^{-24}$ & $4.64\times10^{-25}$ & $2.0\times10^{-26}$ & $2.0\times10^{-24}$ \\ \cmidrule(lr{1em}){3-3}
        
        & & SFTS ** $(\varepsilon^T)$ & {\includegraphics[scale=0.055]{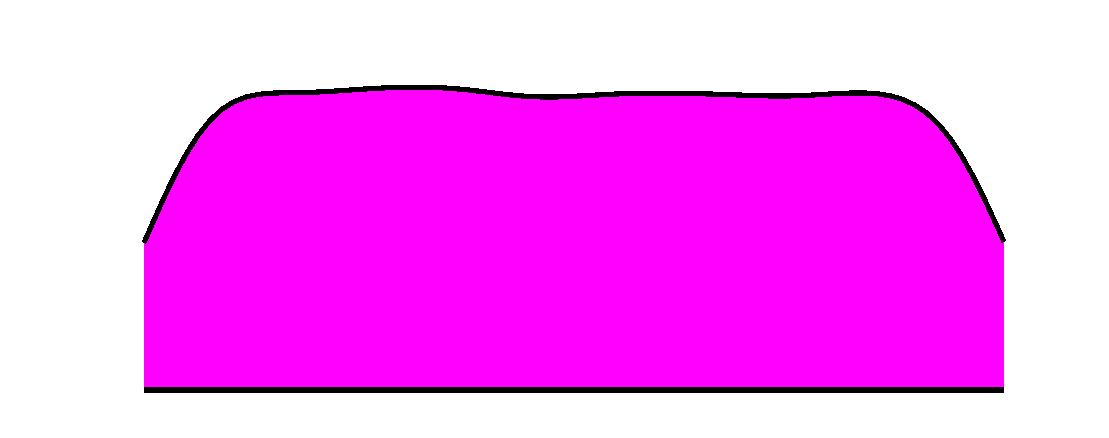}} & - & $-3.1\times10^{-5}$ & -4.28 & -0.02 & +0.02 \\    
        & & $C_{11}$ $Mg_2Sn$   & {\includegraphics[scale=0.055]{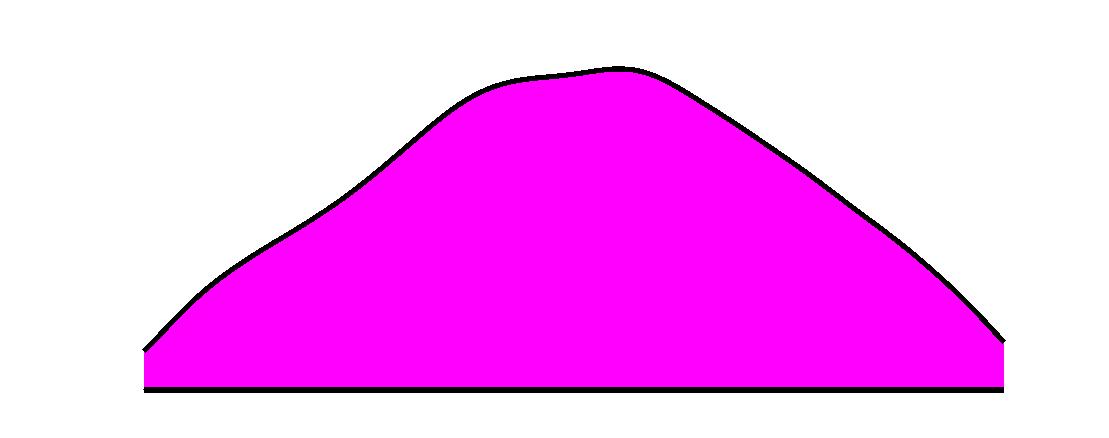}} & $GPa$ & 76.56 & 0.20 & 68.30 & 83.71 \\   
        & & $C_{12}$ $Mg_2Sn$   & {\includegraphics[scale=0.055]{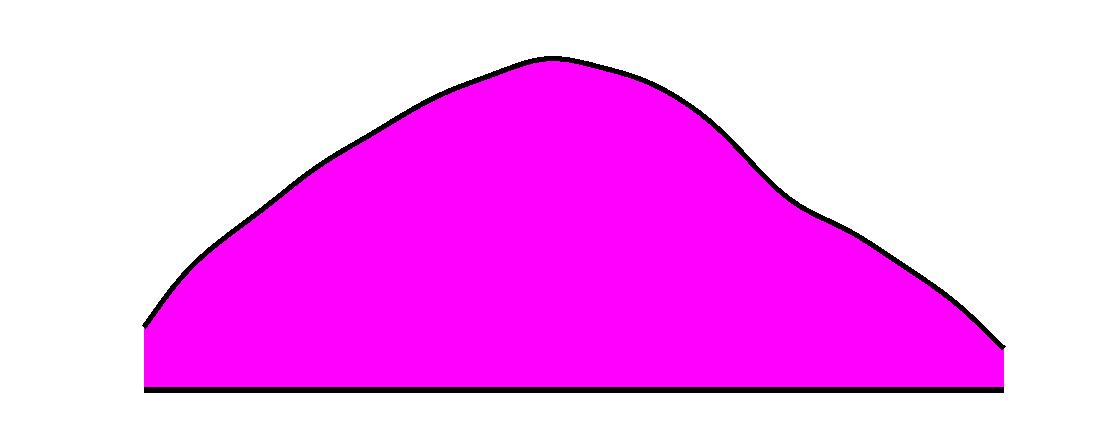}} & $GPa$ & 27.75 & 1.04 & 17.68 & 39.79 \\   
        & \parbox[l]{20mm}{\multirow{1}{*}{\rotatebox[origin=c]{90}{\normalsize{Microelasticity Model Parameters}}}} & $C_{44}$ $Mg_2Sn$   & {\includegraphics[scale=0.055]{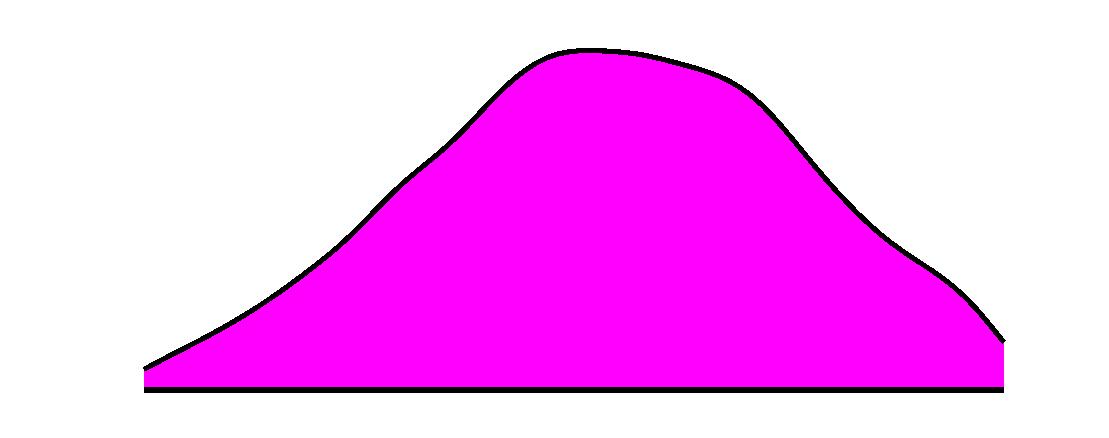}} & $GPa$ & 29.94 & 1.26 & 16.03 & 41.94 \\   
        & & $C_{11}$ $Mg_2Si$   & {\includegraphics[scale=0.055]{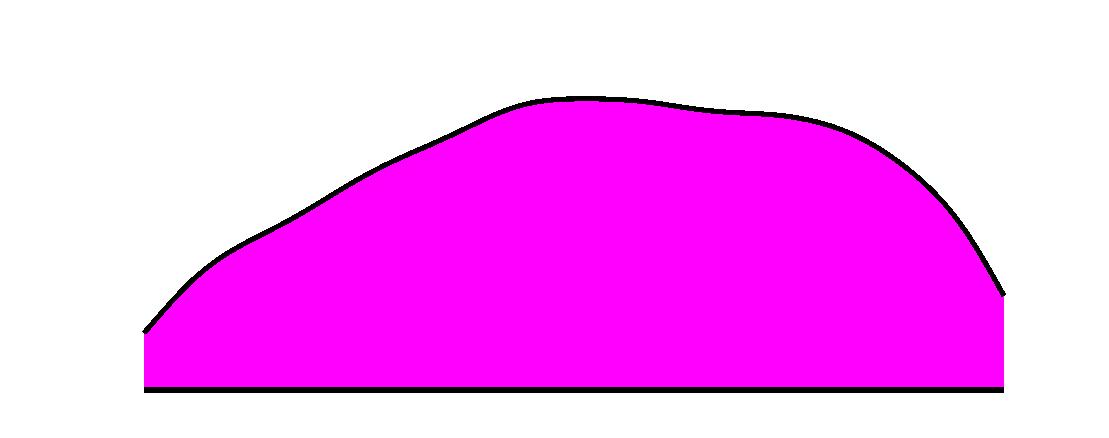}} & $GPa$ & 120.15 & 0.07 & 114.07 & 126.00 \\
        & & $C_{12}$ $Mg_2Si$   & {\includegraphics[scale=0.055]{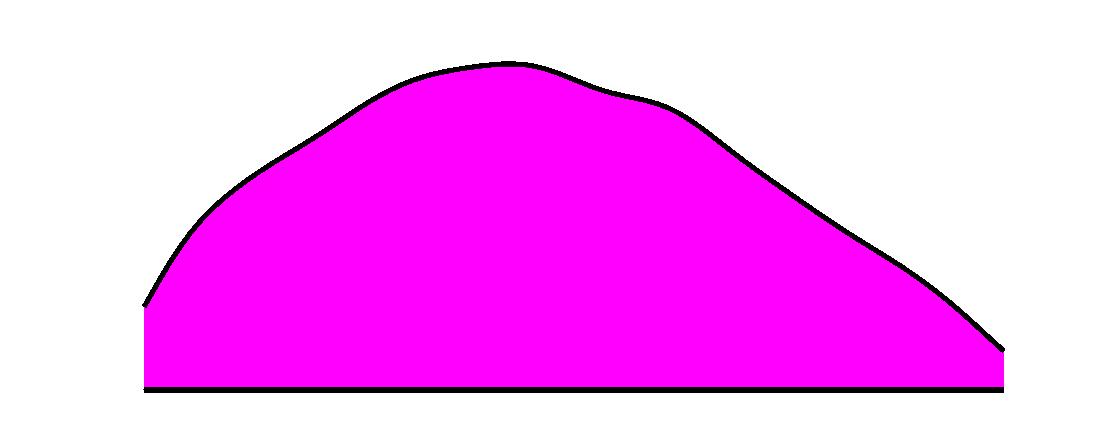}} & $GPa$ & 22.62  & 0.10 & 19.56 & 26.00 \\    
        & & $C_{44}$ $Mg_2Si$   & {\includegraphics[scale=0.055]{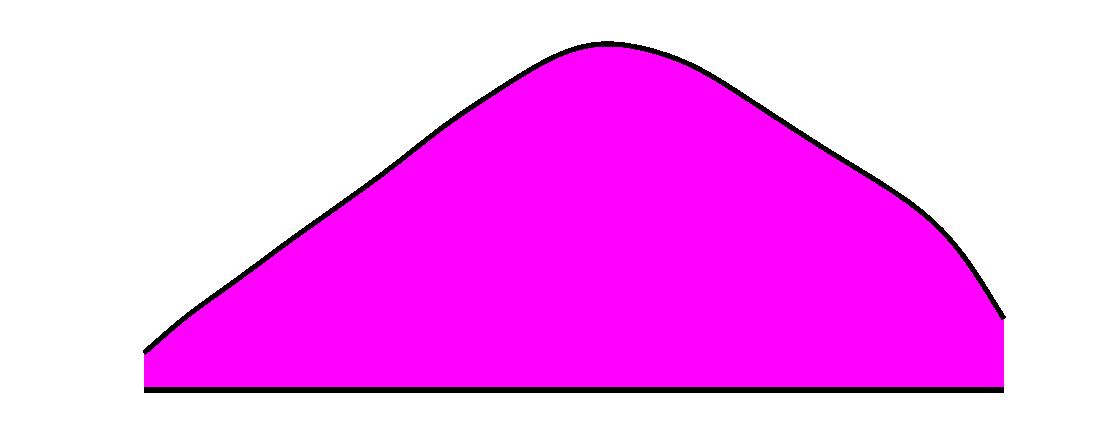}} & $GPa$ & 46.79 & 0.63 & 33.32 & 58.20 \\  
        & & Molar volume $(V_m^{Mg_2Si})$ & {\includegraphics[scale=0.055]{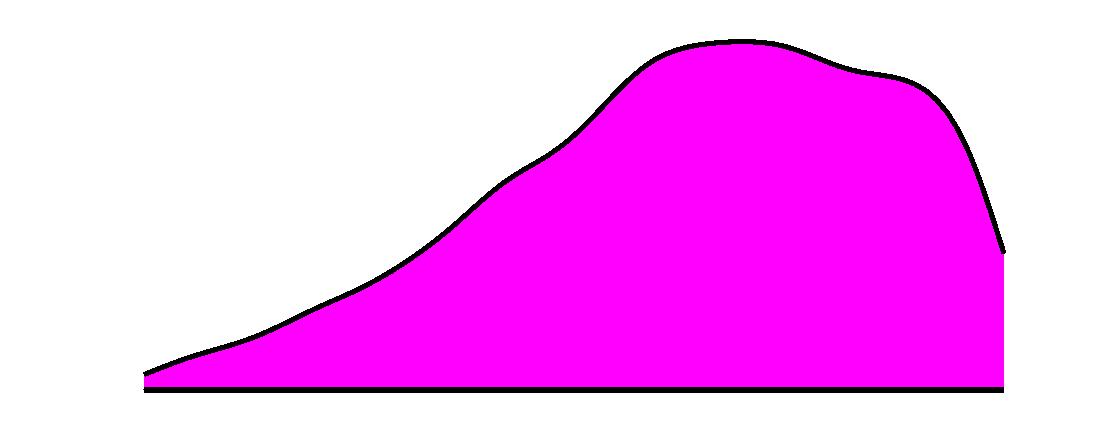}} & $m^{3}.mol^{-1}$ & $5.78\times10^{-5}$ & $2.40\times10^{-7}$ & $4.73\times10^{-5}$ & $6.38\times10^{-5}$ \\    
        & & Molar volume $(V_m^{Mg_2Sn})$ & {\includegraphics[scale=0.055]{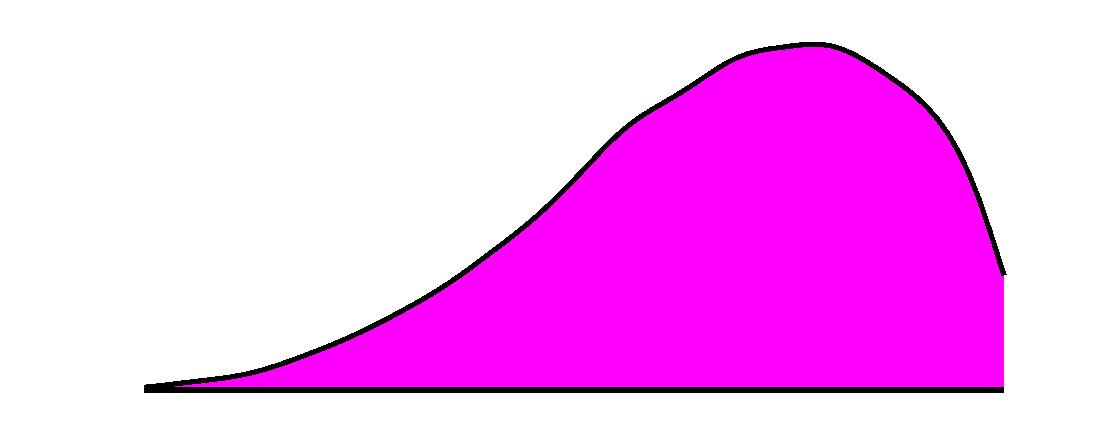}} & $m^{3}.mol^{-1}$ & $4.88\times10^{-5}$ & $1.55\times10^{-7}$ & $3.95\times10^{-5}$ & $5.33\times10^{-5}$ \\       \bottomrule 
    \end{tabular}                                       
    \label{table:1st_order_elastic_cts}                          
    \end{table*} 

\subsection{Uncertainty Quantification/Propagation}\label{3.3}
    
The quantification of uncertainties associated with model parameters and predictions is one of the most important tasks in computational aided materials design. In this regard, quantifying the uncertainties of the model parameters given data is an inverse problem, known as inverse UQ, while obtaining the uncertainties of the predictions of either an individual model or a chain of models through the propagation of the parameter uncertainties is a forward problem, known as forward UP.
    
Generally, UQ of model parameters can be performed in the context of either frequentist or Bayesian inference. In this work, the latter has been applied to probabilistically calibrate the relevant model parameters. Here, the parameter calibration or uncertainty quantification is performed based on Bayes' rule where the parameter posterior probability given data is proportional to the parameter prior probability times likelihood, i.e. $P(\theta|D) \propto P(\theta)P(D|\theta)$ assuming $\theta$ and $D$ are the parameters and data, respectively. Here, the prior probability distribution of the model parameters---which is defined through the available prior knowledge about the parameters---is updated to their posterior probability distribution given data using the likelihood distribution---which is a conditional probability of data given the model parameters. Therefore, the Bayesian inference results from the combination of prior beliefs and likelihood rather than the inference just based on likelihood, as is the case in frequentist approaches. 

The resulting posterior distribution represents the plausible optimal value of the parameters and their uncertainties that can be propagated to the model predictions through different analytical or numerical approaches. To find the statistical properties (mean value and covariance matrix) of the posterior probability distribution, some intractable integrals should be solved; however, these integrations are very difficult to carry out through the conventional numerical approaches. For this reason, Monte Carlo (MC)-based approaches are often used to estimate these integrals \cite{gelman2013bayesian,foreman2013emcee}. In order to obtain well-converged estimates for the relevant integrals, a substantial number of evaluations of the model must be made. Generally, Markov Chain Monte Carlo (MCMC) approaches are the most common used sampling techniques in the context of Bayesian inference. Among these techniques, the Metropolis-Hastings (MH) algorithm has been selected in this work, as described in detail in section \ref{3.3.1}. 
    
We note that it is also important to propagate uncertainties from the parameters to the model predictions since these probabilistic predictions of the given system are the quantities of interest in materials design.

Three commonly used approaches for UP are the analytical methods, the numerical Monte Carlo (MC) methods, and stochastic finite elements. In analytical methods, the output/prediction uncertainties are represented explicitly as functions of parameter uncertainties. These methods are useful when only small ranges of uncertainty are considered, and cannot always be applied to complex, nonlinear models with large parameter uncertainties. Monte Carlo methods involve the estimation of probability density functions (PDFs) for selected model outputs by performing a sufficiently large number of model runs with randomly sampled parameters. For computationally intensive models, the time and resources required for Monte Carlo methods could be  prohibitively expensive. A degree of computational efficiency is accomplished by the use of Modified Monte Carlo (MMC) methods that sample from the parameter distribution in an efficient manner, so that the number of necessary solutions are significantly reduced compared to the simple Monte Carlo method. Such methods include the Fourier Amplitude Sensitivity Test (FAST) \cite{mcrae1982global} and Latin Hyper-cube Sampling \cite{huntington1998improvements}. However, even these require a substantial number of model runs to obtain a good approximation of the output PDFs, especially for cases involving a large number of uncertain parameters. Therefore, there is a need for even more computationally efficient ways for propagating uncertainty in complex/expensive models and this is currently being investigated by the present authors. 

In this work, a forward UP analysis based on optimal sampling methods has been applied to propagate the uncertainties across the chain of models, which is explained in section \ref{3.3.2}.
    
    
    \subsubsection{Applied MCMC Approach for the Parameter Uncertainty Quantification}
    \label{3.3.1}
    
The MCMC-MH toolbox in Matlab has been utilized to calibrate the Gibbs free energy parameters of a CALPHAD model for the pseudo-binary Mg$_2$(Si$_x$Sn$_{1-x}$) system. Since there is no prior knowledge about the distribution of the parameters and their correlations, a uniform prior probability distribution has been considered for the model parameters in this work. However, the initial values ($\theta^0$) and ranges of the parameters have been determined based on the deterministically calibrated parameters in Thermo-Calc software in order to make the parameter convergence faster during MCMC sampling. After defining the prior probability distribution, the parameters (candidates) are randomly sampled from an arbitrary posterior proposal distribution ($q$) iteratively. In this work, the proposal distribution is an adaptive multi-variate Gaussian distribution that is centered at the last accepted parameter vector in the MCMC chain with a covariance matrix adapting during MCMC sampling based on the resulting covariance from the previous parameters in the chain, according to Haario et al \cite{haario2001adaptive}. In each iteration, the sampled candidate is accepted or rejected based on the Metropolis-Hastings ratio:  

    \begin{equation}
    \label{eq 14}
    M-H=\frac{p(\theta^{cand})p(D|\theta^{cand})}{p(\theta^{z-1})p(D|\theta^{z-1})} \frac{q(\theta^{z-1}|\theta^{cand})}{q(\theta^{cand}|\theta^{z-1})}
    \end{equation}
    
This ratio compares the posterior probability of the sampled candidate ($\theta^{cand}$) with its counterpart for the last accepted parameter vector in the MCMC chain ($\theta^{z-1}$) through the metropolis ratio (the first ratio in Equation \ref{eq 14}), and also compares the probability of moving from $\theta^{z-1}$ to $\theta^{cand}$ with the probability of the reverse move through the Hastings ratio (the second ratio in Equation \ref{eq 14}) in the case that the posterior proposal distribution is asymmetric. It should be noted that the likelihood function is also a Gaussian distribution centered at the given data ($D$) with a variance that corresponds to the uncertainty of the data. Since the uncertainty of the data used in this work is not clear, an unknown variance has been considered for the likelihood which has been updated as a hyper-parameter with the model parameters during MCMC sampling, which has been explained in more detail by Gelman et al. \cite{gelman2013bayesian}.  Min(M-H,1) indicates the acceptance probability of the candidate in each iteration. $\theta^{z}$ equals $\theta^{cand}$ in the case of acceptance which updates the mean value of the proposal distribution; while it equals $\theta^{z-1}$ if the candidate is rejected. The sampling process proceeds until the convergence of MCMC chain to a stationary distribution, which represents the parameter convergence during MCMC process. After discarding the burn-in period which is the initial MCMC samples before the parameter convergence, the mean values of the remaining samples and the square root of the diagonal terms in their covariance matrix represent the plausible optimal values and uncertainties of the parameters, respectively.
    
\subsubsection{Applied Uncertainty Propagation Approach}
\label{3.3.2}

As mentioned earlier, a forward UP analysis has been considered in this work to propagate uncertainties from thermodynamic parameters to Gibbs free energy of phases to microstructural characteristics in Mg$_2$(Si$_x$Sn$_{1-x}$) system. In this regard, a group of sampled parameter vectors has been considered as representative of the multi-variate posterior distribution of the thermodynamic parameters and run through the applied CALPHAD model to find the corresponding responses for Gibbs free energy of phases and phase diagram. To find 95\% Bayesian credible intervals (BCI), 2.5\% of the samples from both the lower and upper bounds of the response distributions have been removed. The same approach has been used to propagate uncertainties from thermodynamic, microelastic and kinetic parameters to microstructural characteristics throughout a phase field model. However, a Gaussian copula approach has been used to sample a reasonable number of parameter vectors from the distributions of microelastic and kinetic parameters---assuming they are independent---as well as the multi-variate posterior distribution of thermodynamic parameters obtained by the MCMC technique.

\subsubsection{Sampling Methodology}\label{sec:Sampling_method}

Here we note that propagating uncertainties from high-dimensional input sets has to be carried out in a way that minimizes the number of samples ( a typical naive sampling scheme using MCMC approaches may require $\mathcal{O}\left(1,000,000\right)$ random samples) while at the same time accounting for the statistical correlations among input parameters---for example, parameters in CALPHAD thermodynamic descriptions tend to be highly correlated. 

To construct sample sets with correct marginal distributions and preserved pairwise correlations, we instead make use of Gaussian copulas. A copula is a function that relates the joint cumulative distribution function (CDF) of multiple variables to their marginal CDFs and their correlations~\cite{choi2010reliability}. To begin, we assume that we have available marginal distributions, $f_{X_i}(x_i)$, for each parameter, where $X_i$ denotes the random variable associated with the $i^{th}$ parameter, $i \in \{1, 2, \ldots, K\}$, and $x_i$ is a particular realization of $X_i$. We also have available pairwise correlation coefficients, $\rho_{i,j}$, where $i,j \in \{1,2,\ldots,K\}$, and 
    \begin{equation}
    \label{eq 15}
        \rho_{i,j} = \frac{\textrm{Cov}(X_i,X_j)}{\sigma_{X_i}\sigma_{X_j}},
    \end{equation}

where $\textrm{Cov}$ denotes the covariance and $\sigma$ denotes the standard deviation.  These correlation coefficients are stored in a matrix, $R \in [-1,1]^{K\times K}$, which we use to create the proper correlation structure among the pairwise joint distributions.  To do this, we create a set of independent, identically distributed random vectors, $\mathbf{G}_1, \mathbf{G}_2, \ldots, \mathbf{G}_n \;\; \textrm{i.i.d.} \sim \mathcal{N}(\mathbf{0},R)$, thus, each $G_j$, $j \in \{1, 2, \ldots, n\}$ is a $K$-dimensional random vector with a zero mean multivariate normal distribution with covariance, $R$.  From this set of random vectors we can create a sample set, $\mathbf{g}_1, \mathbf{g}_2, \ldots, \mathbf{g}_n$, of $n$ samples from $\mathcal{N}(\mathbf{0},R)$, where $\mathbf{g}_j$ are realizations from each identically distributed $\mathbf{G}_j$.  Thus, we have $n$ samples from standard normal distributions marginally, and each pairwise joint density has the desired correlation in the sample set.  From this set of samples, we generate uniformly distributed samples from $\{\mathbf{u}_j = (\Phi(g_j^1), \Phi(g_j^2),\ldots,\Phi(g_j^K))\}_{j=1}^n$, where $\Phi$ is the cumulative distribution function of a standard normal random variable, and $\mathbf{g}_j = (g_j^1, g_j^2, \ldots, g_j^K)$.  This results in $n$ samples from a vector of uniformly distributed marginal random variables with our desired pairwise correlations preserved.  The final step makes use of the inverse cumulative marginal distributions of each of our parameters, $F_{X_i}^{-1}(x_i)$.  From these inverse cumulative marginal distributions and the samples $\mathbf{u}_j$, we compute $\{ \mathbf{x}_j = (F_{X_1}^{-1}(u_j^1), F_{X_2}^{-1}(u_j^2),\ldots,F_{X_K}^{-1}(u_j^K)) \}_{j=1}^n$, where $\mathbf{u}_j = (u_j^1, u_j^2, \ldots, u_j^K)$, which is a set of $n$ sample vectors sampled from the correct marginal distributions provided at the outset and preserved pairwise correlation information.
    
\begin{figure*}[!ht]
        \centering
        \begin{tabular}{ccc}
            \includegraphics[width=.30\textwidth]{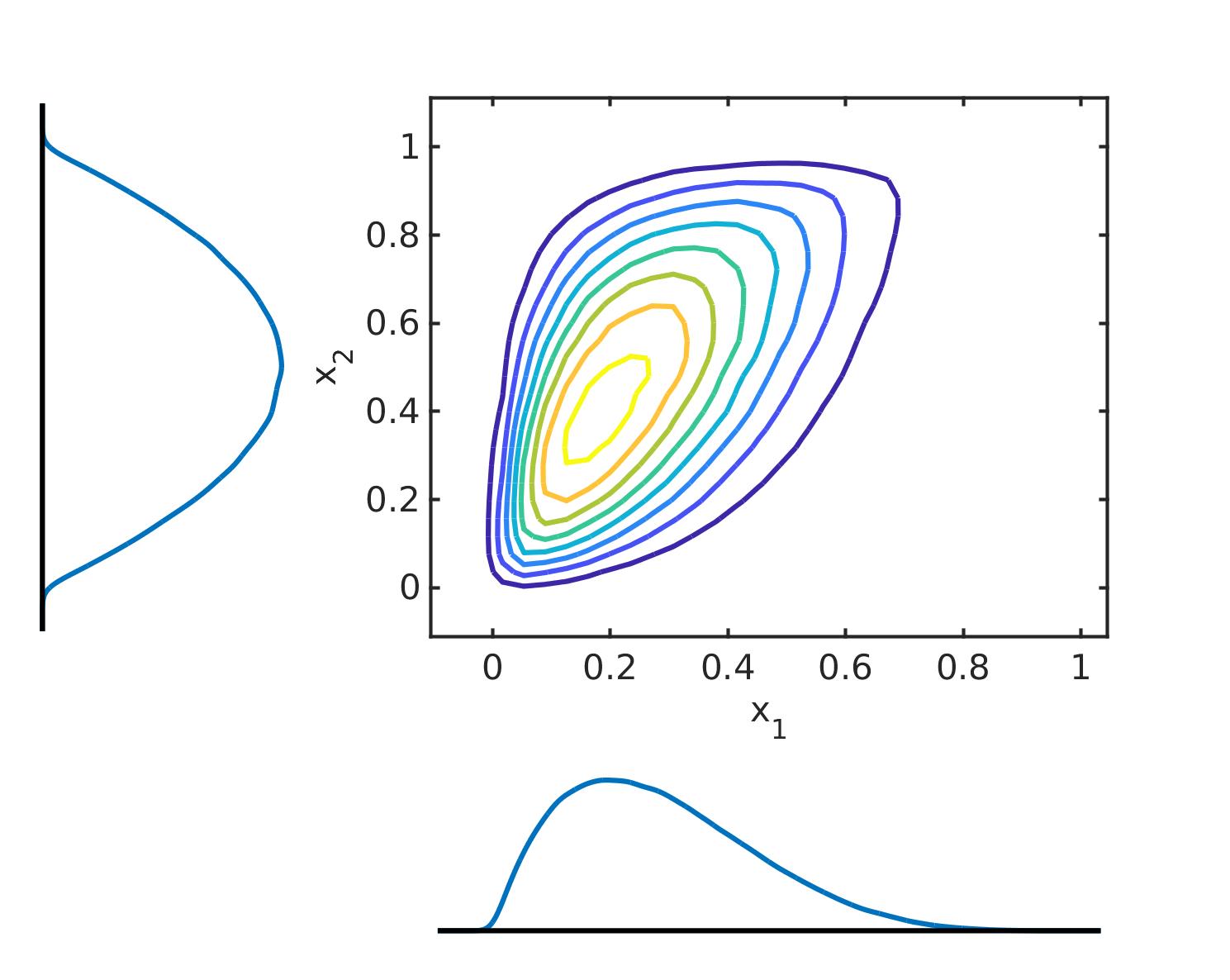} & \raisebox{2.0cm}{\Huge{$\approx$}} &         \includegraphics[width=.30\textwidth]{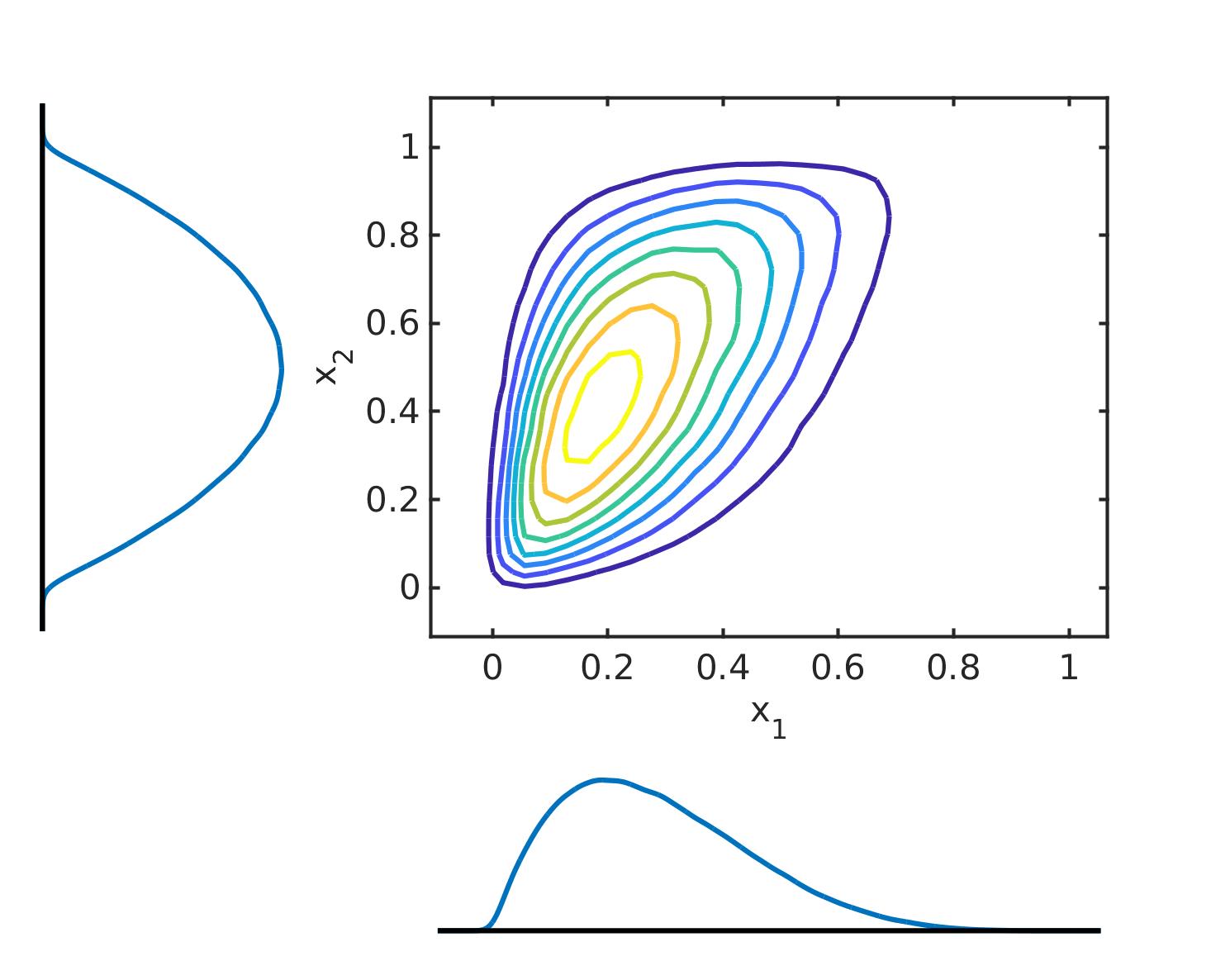}\\    
            $\left[f_{X_1},X_2(x_1,x_2),f_{X_1}(x_1),f_{X_2}(x_2),R  \right]$  & &  
            $\left[f_{X_1},X_2(x_1,x_2),f_{X_1}(x_1),f_{X_2}(x_2),R  \right]$      \\
            \begin{tikzpicture}
            \coordinate (a) at (0,1.2);
            \coordinate (b) at (0,0);
             \draw[->, >=latex, blue!25!white, line width=8pt]   (a) to node[black]{$\mathcal{N}(0,R)$} (b) ;
            \end{tikzpicture}
            & &
            \begin{tikzpicture}
            \coordinate (a) at (0,0);
            \coordinate (b) at (0,1.2);
            \draw[->, >=latex, blue!25!white, line width=8pt]   (a) to node[black]{$F^{-1}_{X_i}(u_j^i)$} (b) ;
            \end{tikzpicture}  \\        
            $\left[f_{G_1}(g_1),f_{G_2}(g_2)  \right]$  & &  
            $\left[f_{U_1}(u_1),f_{U_2}(u_1),R  \right]$      \\       
            \includegraphics[width=.30\textwidth]{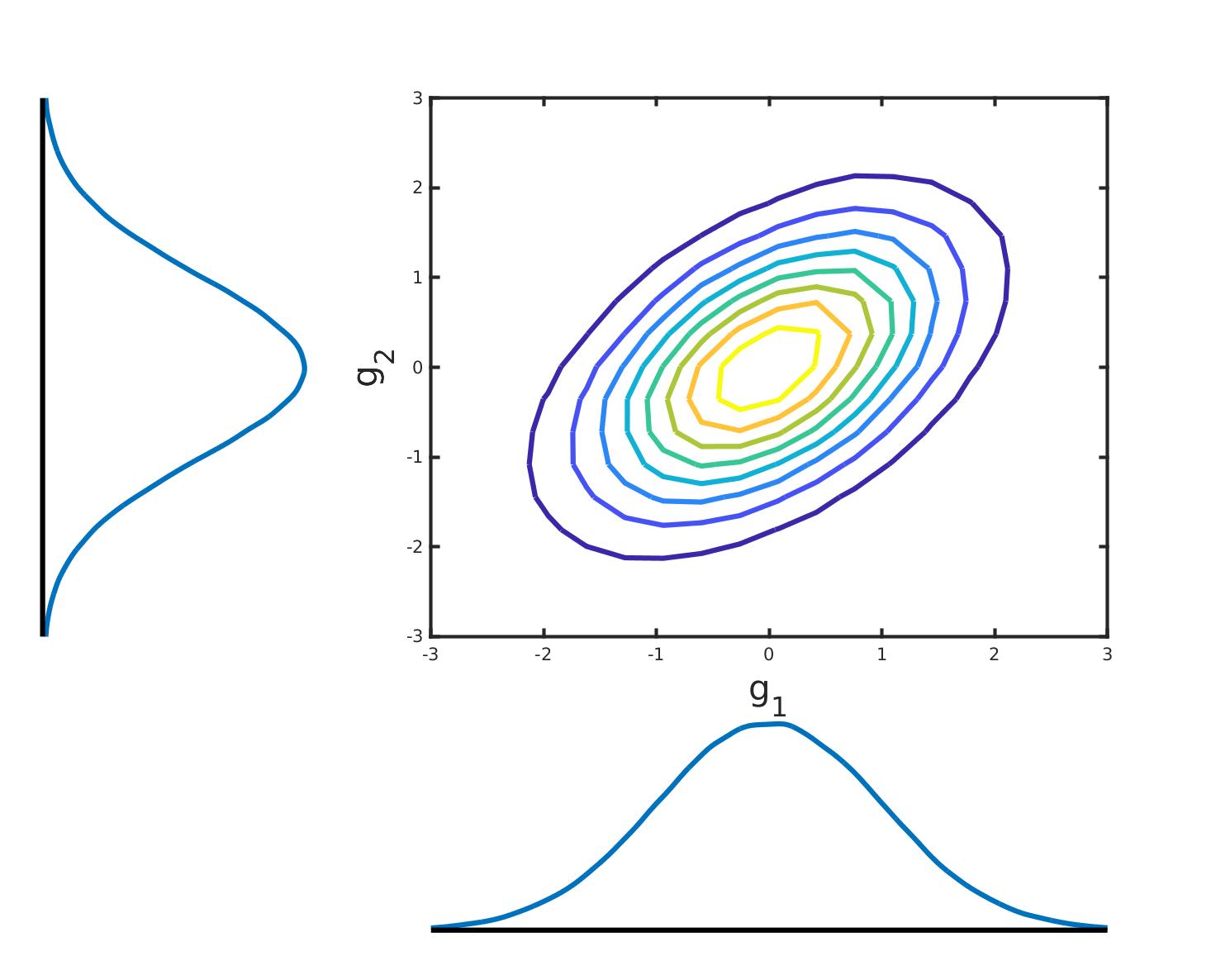} &
            \raisebox{2.00cm}{
            \begin{tikzpicture}
            \coordinate (a) at (0,0);
            \coordinate (b) at (2,0);
            \draw[->, >=latex, blue!25!white, line width=10pt]   (a) to node[black]{\large{$\Phi$}} (b) ;
            \end{tikzpicture}} &
            \includegraphics[width=.30\textwidth]{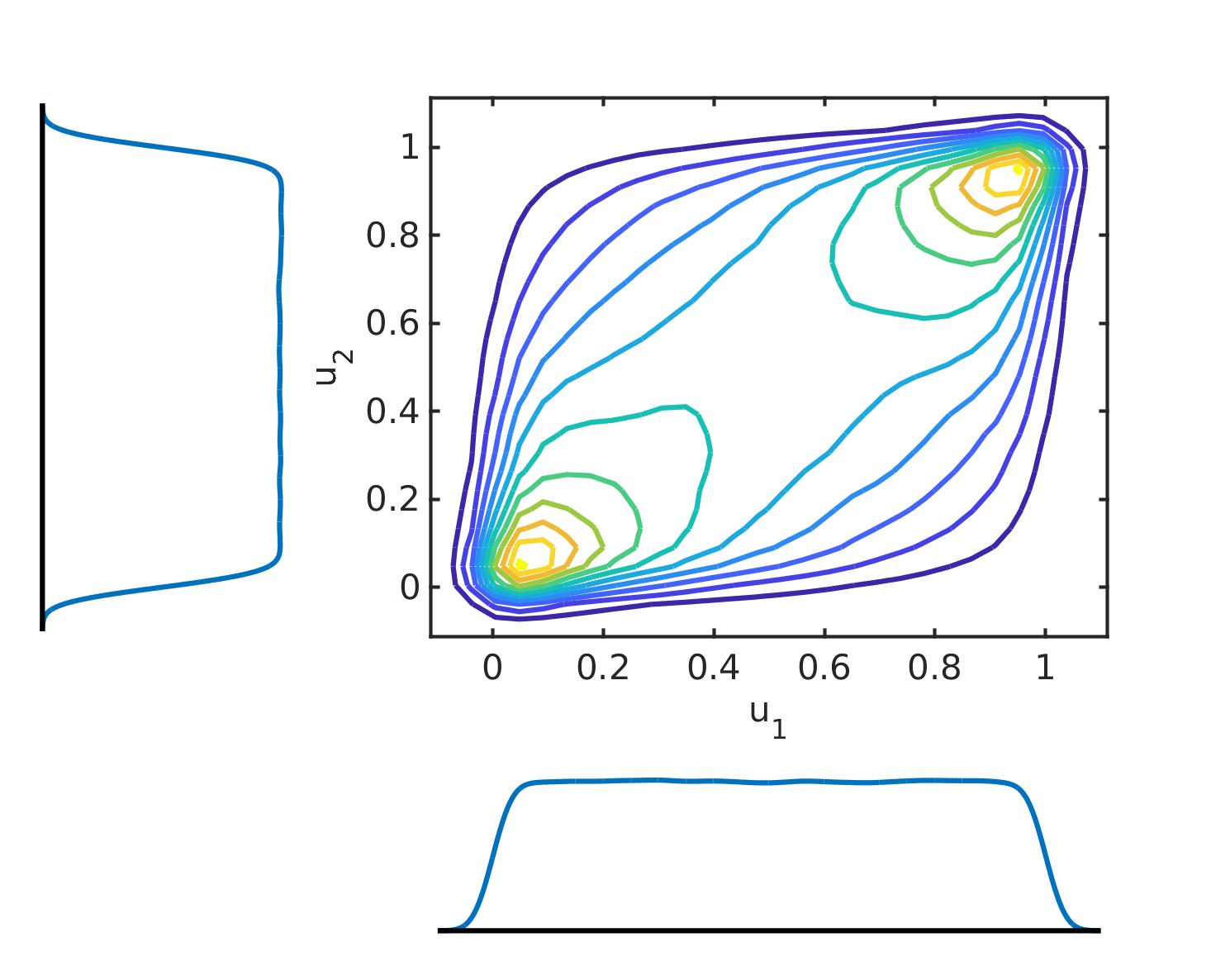} \\
            \end{tabular}
            \caption{Two-Dimensional demonstration of our sampling strategy that preserves marginal distributions and pairwise correlations.}
            
        \label{fig 2}
    \end{figure*}
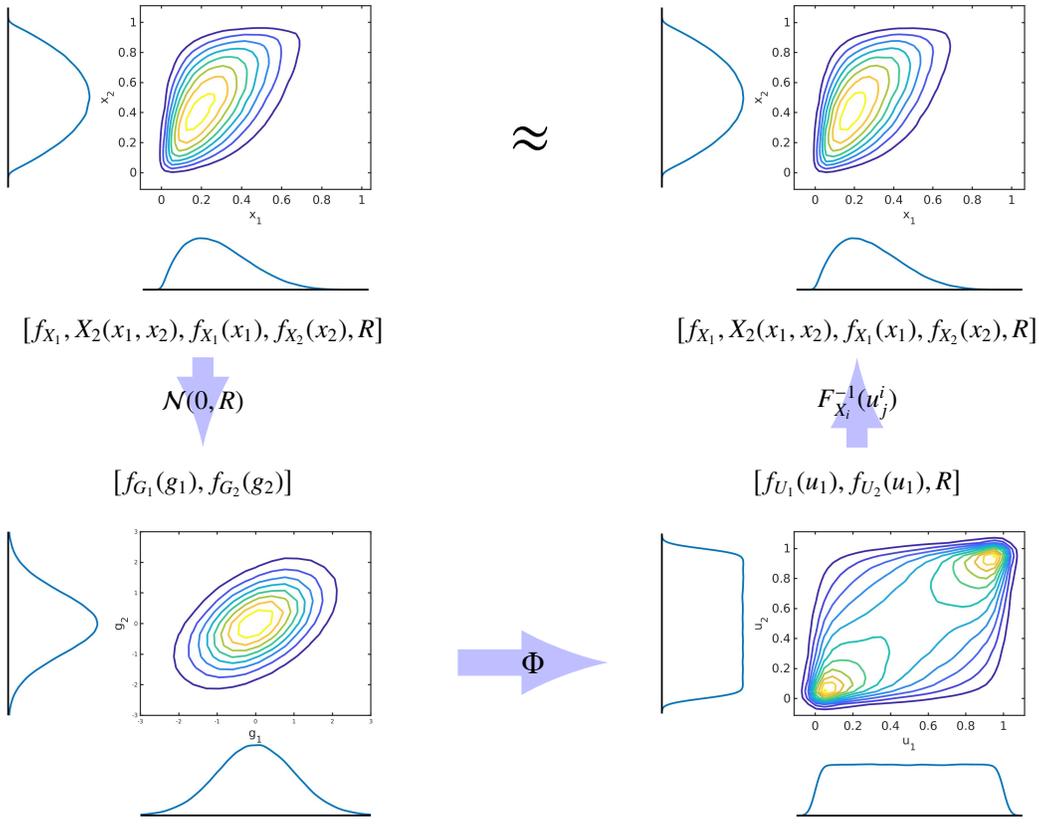

Our sampling methodology is demonstrated here for a two-dimensional random vector in Figure~\ref{fig 2} to make the preceding discussion more concrete.  The top left plot of the figure is the original two-dimensional joint distribution with each marginal shown as well.  This information is distilled into the marginal distributions, $f_{X_1}(x_1)$ and $f_{X_2}(x_2)$, as well as the correlation coefficient matrix, $R$.  From this correlation information, two correlated standard normal marginal distributions are created and shown in the bottom left plot of the figure.  Samples from these distributions are passed individually through the standard normal cumulative distribution, $\Phi$, which leads to two correlated uniform random variables shown in the bottom right plot.  Samples from these distributions are passed individually through the inverse marginal distributions of each parameter, respectively.  This results in samples from correct marginal distributions with preserved pairwise correlation.  This can be seen by comparing the top left and top right plots.  The top left plot is the true joint distribution and the top right plot is the joint distribution generated via this methodology, which is correct in terms of the marginal distributions and the correlation coefficient between $X_1$ and $X_2$.

\section{Results and Discussion}\label{sec:results_Discussion}
    
Solid state reactions are not only influenced by chemical driving forces, but also by the interfacial and elastic energy contributions. The bulk free energy ($f^{bulk}$) greatly dominates the overall phase stability of a system but strong couplings to elastic fields arising from lattice and elastic mismatch as well as anisotropy in the properties play a dominant role in controlling the overall thermodynamic stability of the system as well as the topology and morphology of the resulting microstructures. Given the influence of bulk thermodynamic properties, the uncertainties in the bulk free energy require precise quantification first. This will be described in section \ref{sec:4.1}. Then the quantified uncertainties of the bulk free energy, the kinetic and microelastic parameters are propagated to microstructural characteristics in section \ref{4.2} through the elastochemical phase field model.
     

    
    
\subsection{Uncertainty Quantification of Gibbs free energy parameters and Phase Diagram}\label{sec:4.1}

A thorough parameter uncertainty analysis of the CALPHAD model for the Mg$_2$(Si$_x$Sn$_{1-x}$) pseudo-binary system has been performed through the MCMC technique (explained in section \ref{3.3.1}) against the calculated composition-temperature data sampled from the phase diagram proposed by Kozlov et al. \cite{kozlov2011phase}. Then, the calculated parameter uncertainties have been propagated to the Gibbs free energy of phases and phase diagram. 

As shown in Table \ref{table:1st_order_elastic_cts}, three parameters $^{0}a$, $^{0}b$ and $^{1}a$ are selected for each phase in the CALPHAD model, i.e. Mg$_2$X\{Sn,Si\} and liquid (six parameters in total). As mentioned earlier, the deterministically optimized parameters obtained from Thermo-Calc have been utilized as initial parameter values for MCMC sampling process; however, lack of knowledge about the parameter probability distributions resulted in the consideration of a uniform (non-informative) prior distributions for the parameters. In addition, $\pm 50\%$ of the parameter initial values have been considered as the parameter ranges during this process.

\begin{figure}[h!]
        \centering
        \includegraphics[scale=0.3]{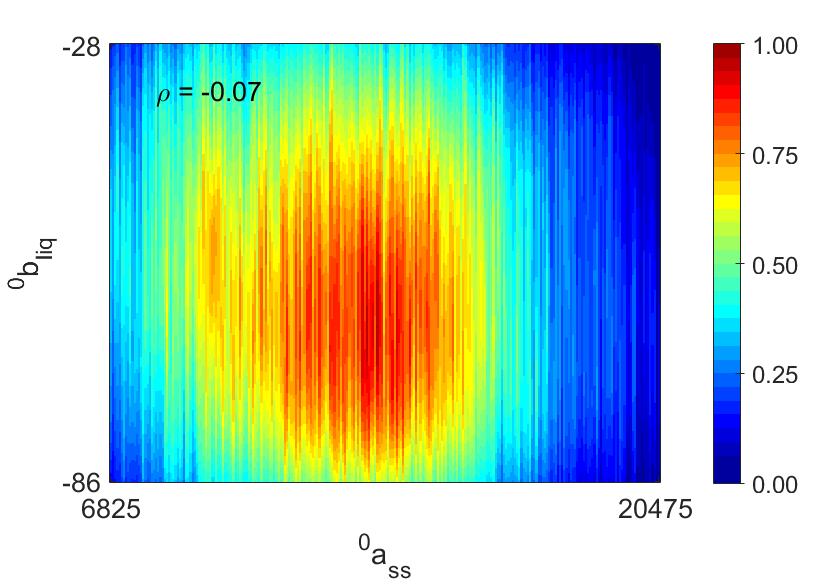}
        \caption{Joint frequency distribution between a selected pair of parameters.}
        \label{fig 3}
    \end{figure}

During MCMC calibration, 100,000 samples have been generated to ensure parameter convergence. In this regard, plotting the joint frequency distribution of each pair of the model parameters can graphically show the parameter convergences in the parameter space. For example, one of these plots has been shown in Figure \ref{fig 3}. As observed in this figure, the red region with the highest density of parameter samples indicates the convergence region in the pair parameter space. Moreover, these kinds of plots can qualitatively show a linear correlation between the model parameters based on the linearity and direction of the convergence region.
    
The degree of correlation between each parameter pair can also be quantified through the Pearson linear coefficient \cite{battle_2004} denoted by $\rho_{i,j}$ in Equation \ref{eq 15}. It is worth noting that the linear coefficient is a quantity between -1 and 1. Generally, parameters are uncorrelated if $\rho$ is close to 0, but highly correlated if $\rho$ is close to either -1 or 1. In addition, the negative and positive signs indicate the correlation direction. In Figure \ref{fig 3}, a semi-circular shape of convergence region and a very close value of $\rho$ to zero imply negligible correlation between $^{0}a_{ss}$ and $^{0}b_{liq}$.

    

\begin{figure}[!ht]
        \centering
        \includegraphics[width=0.45\textwidth]{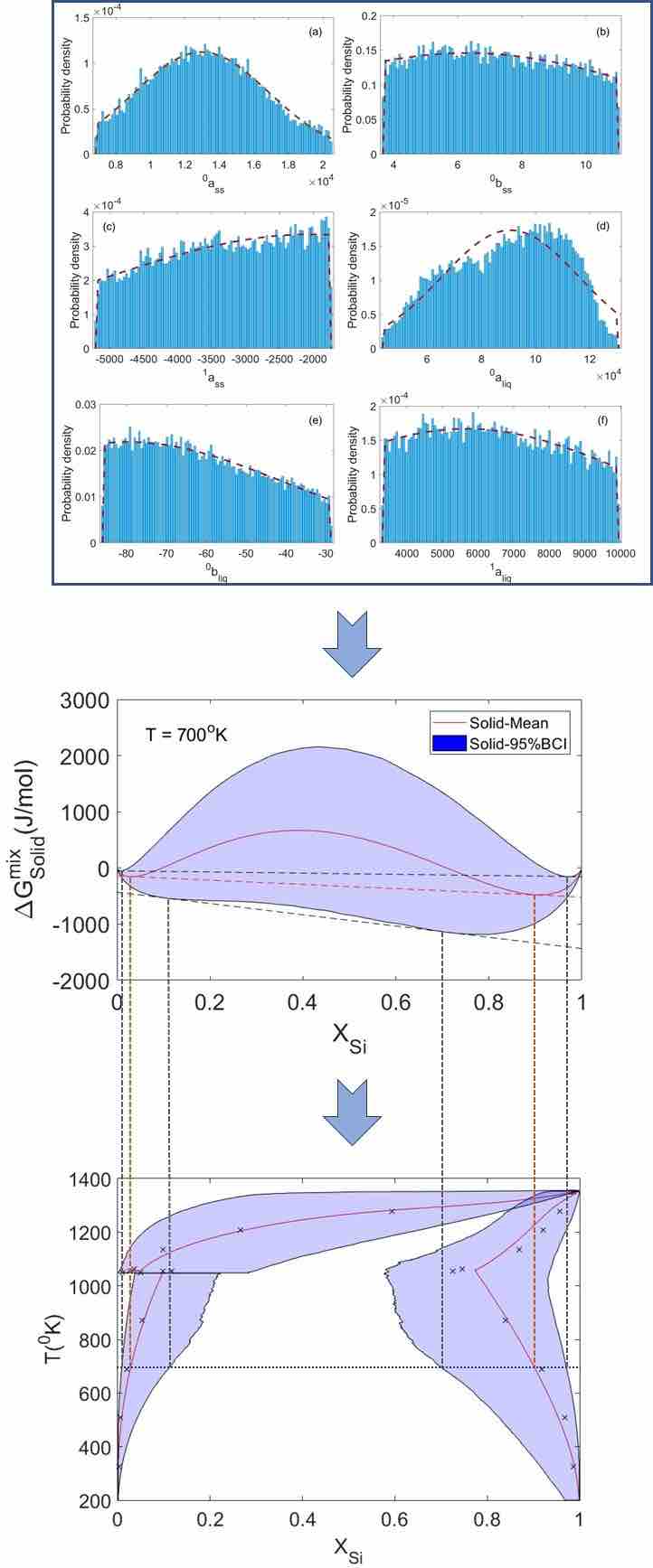}
        \caption{An illustration of UP from thermodynamic parameters to the curves associated with the molar Gibbs energy of mixing to the phase diagram for Mg$_2$X\{Sn,Si\} quasi-binary system at 700 $^\text{o}$K.}
        \label{fig 4}
    \end{figure}

After discarding the burn-in period in the beginning of the MCMC sampling, the marginal probability distribution of each model parameter can be plotted as shown with blue color in Table \ref{table:1st_order_elastic_cts} and Figure~\ref{fig 4}. As observed in these distributions, all parameters show distributions close to truncated normal. Therefore, a truncated normal probability distribution has been fitted to each marginal probability distribution, where the hyper-parameters ($\mu$ and $\sigma$) of these truncated distributions have been estimated through a maximum likelihood method. The estimated $\mu$ and $\sigma$ of each parameter probability distribution has also been listed in Table \ref{table:1st_order_elastic_cts}, which represent the plausible optimal values and uncertainties of the model parameters.

The parameter uncertainties in Table~\ref{table:1st_order_elastic_cts} have been propagated to the molar Gibbs free energy of mixing for different phases in the system to the phase diagram. As mentioned in section \ref{3.3.2}, the last 5,000 MCMC samples as an ensemble of the whole convergence region have been used in a forward model analysis scheme to identify the variation of the lines in the phase diagram. Then, 95\% BCIs have been determined by discarding 2.5\% of the resulting samples from the above and below the variation band at any specified Si composition in the Gibbs energy of mixing curves or the phase diagram. In this regard, Figure~\ref{fig 4} demonstrates uncertainty propagation across different levels of CALPHAD modeling. Here, just an example of molar Gibbs energy of mixing for solid phases (Mg$_2$X\{Sn,Si\}) at 700 $^\text{o}$K has been plotted to show how uncertainties propagate from the Gibbs free energy curves to the phase diagram. This analysis has been repeated for different temperatures in the range from 200 to 1400 $^\text{o}$K to construct the whole phase diagram and its uncertainties. Based on Figure~\ref{fig 4}, it should be noted that there are very high uncertainties in the curves for the Gibbs energy of mixing as well as the resulting phase diagram in most temperatures. These high uncertainties imply very high impacts of the Gibbs free energy parameter variations on the phase equilibria in this system.
    
    
    
\subsection{Effect of Parameter Uncertainties on Microstructure}
\label{4.2}

A complete phase-field model for the Mg$_2$(Si$_x$Sn$_{1-x}$) system requires the definition of a large number of parameters---see Table~\ref{table:1st_order_elastic_cts}. The proper characterization of all relevant parameters has been a major challenge in mesoscale microstructural models, and it is a common practice to define the values of these parameters deterministically and ignore actual variabilities. In turn, this might ignore the evidence variability on the model outputs. This implies that common sources of uncertainty may have little or no impact on the selection of the value of model parameters. The uncertainties in parameters of the Gibbs energy, obtained through the MCMC approach, are combined with the prior distribution of the parameters of other sub-models (microelasticity and kinetic) to study the microstructural evolution using the phase-field model. The prior information of the all parameters are reported in Table~\ref{table:1st_order_elastic_cts}. Here we elaborate on the source of these prior distributions.
    

The process of phase separation is influenced by the elastic anisotropy, and the hardness enhancement observed upon the age hardening relies on a shear modulus difference between the formed domains as well as their coherency strain \cite{fried1999coherent}. Thus, it is of primary interest to take into account the uncertainities in the elastic properties of the Mg$_2$(Si$_x$Sn$_{1-x}$) system to study the contributions in microstructural uncertainty.
    
\begin{figure*}[ht!]
        \centering

        \includegraphics[width=\textwidth]{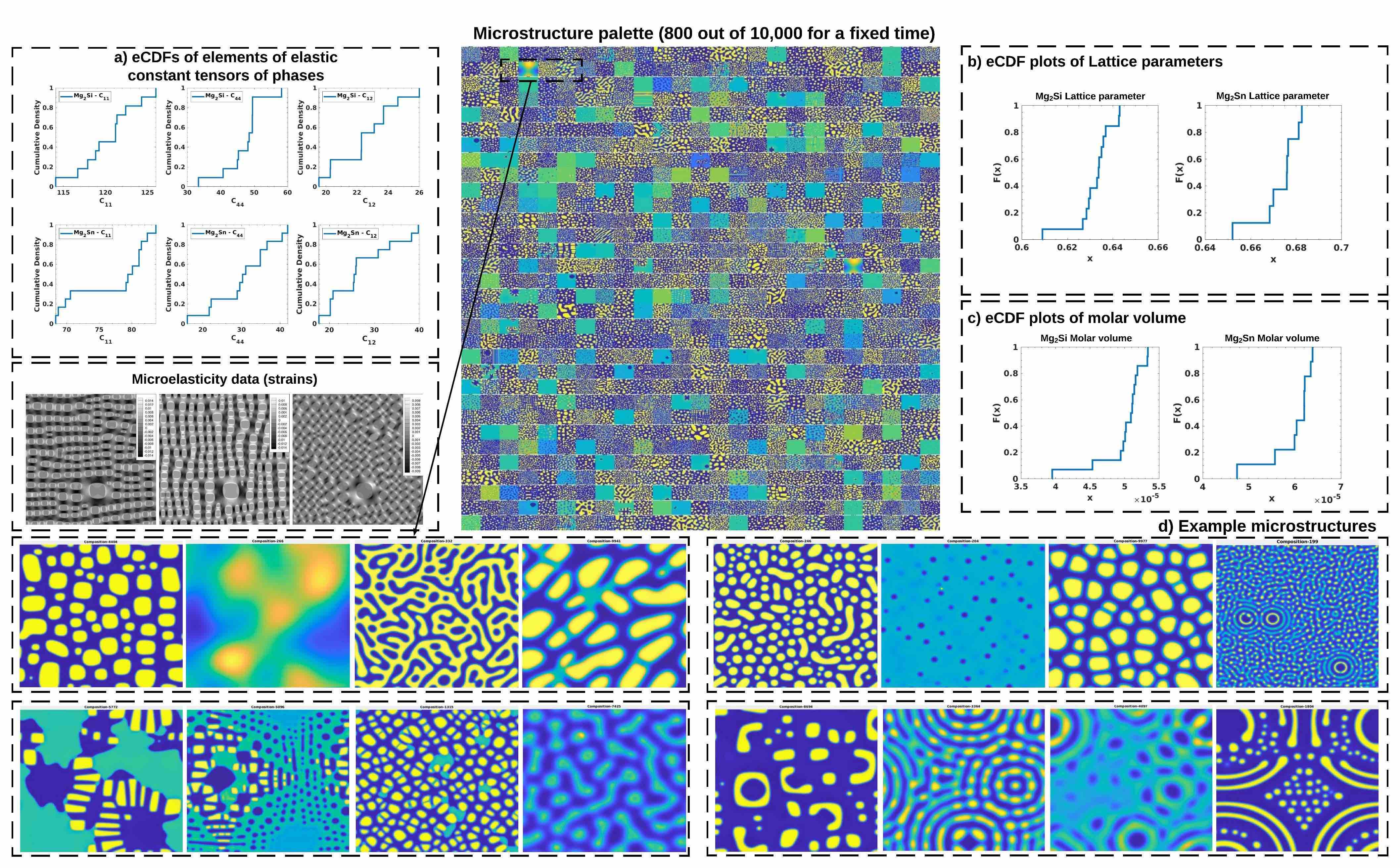}

        \caption{Center) Microstructure library from 800 phase-field runs (out of 10,000 run) for the Mg$_2$X\{Si,Sn\} system at a fixed time. 10,000 points in a 18-dimensional parameter space were sampled from the prior distributions shown in Table \ref{table:1st_order_elastic_cts}. ecdf plots for a) elastic constants, b) lattice parameters, c) molar volumes of the two product phases. d) Example microstructures extracted from the palette.}
        \label{fig:mic_schematic}
    \end{figure*}

The \emph{ab-initio}-based elastic constants (C$_{11}$, C$_{12}$, and C$_{44}$) for (c)-Mg$_2$Sn and (c)-Mg$_2$Si are reported in the literature---we note that these quantities have yet to be measured experimentally. These include both 0$^\circ$K, and high temperature data which are calculated by taking the entropic and/or harmonic effects in the structure. The empirical cumulative density functions (ecdf) of C$_{11}$, C$_{12}$, and C$_{44}$ for both phases are shown in the left side of the microstructure palette in Fig. \ref{fig:mic_schematic}(a). Both \emph{ab-initio} and experimental lattice parameters for the cubic Mg$_2$(Si$_x$Sn$_{1-x}$) system at room/high temperature for different ranges of compositions are considered. These values are either provided for individual phases (Mg$_2$Sn and Mg$_2$Si), or for the parent phase as a function of composition. The ecdf of the lattice parameter data are shown in Fig.~\ref{fig:mic_schematic}(b). Using these data, the range of SFTS ($\varepsilon^{0}_{ij}$) for Mg$_2$Sn and Mg$_2$Si is estimated. This range is used to draw samples from a uniform distribution for propagation of uncertainty in the microelasticity model. 

The molar volumes ($V_m$) of the phases are also obtained based on both calculated and measured data. The calculated data is obtained from $V_m^{cryst} = V_{cell}.N_v/Z$ where $V_{cell}$ is the volume of the unit cell, $N_v$ is the Avogadro's number, and $Z$ is the number of atoms in the cell. Z=3 for both phases considering the fact that the crystal structure of both Mg$_2$Sn and Mg$_2$Si is Fm$\bar{3}$m. The distribution is shown in Fig.~\ref{fig:mic_schematic}(c), which is very similar to the distribution of lattice parameters shown in this figure. Lower bound of this distribution belongs to the data obtained by division of the molar mass and density distributions.

Experimental diffusion data are used to estimate the atomic mobility. The recent combinatorial diffusion couple study by Viv\'es \etal~\cite{vivès2014combinatorial}, and the indirect calculation of inter-diffusion coefficient using the forward-simulation method provides an insight about the diffusion at 600 $^\circ$C. Their calculations suggest an estimated diffusion coefficient of $\approx$ $10^{-15}$ $ms^{-2}$. There is no other information regarding diffusivity measurements to the best of our knowledge. Accordingly, the mobility is estimated by the Einstein's rule ($M=\frac{D}{RT}$), or ($M= \frac{D}{(\frac{\partial^2 f}{\partial c^2})}$) \cite{yi2018strain}. Hence, a uniform sampling is carried out around this value. Initial composition of the alloy is taken to be uniformly distributed between the $x_{Si}=0.3-0.5$ to make sure we are under the chemical miscibility region. 
    
In order to propagate the uncertainties of the prior data, it is necessary to carry out high-throughput phase field simulations of microstructure evolution in the Mg$_2$(Si$_x$Sn$_{1-x}$) system. Using the strategy described in section~\ref{sec:Sampling_method}, we sampled 10,000 combinations of the parameters out of the prior distributions, which have been fed to the phase-field solver (Fortran code). The whole process of 1) data initiation, 2) environment preparation and 3) Fortran code run in the \#Terra super-computing cluster at Texas A\&M University is automated through a Python wrapper. A square simulation cell with 512$\times$512 grid numbers, where $L_x=L_y=350$ nm is used to perform the simulations. The snapshots of the obtained microstructures for 800 samples are shown in the form of a palette in Fig.~\ref{fig:mic_schematic}. Sixteen example microstructures (same evolution time) are also extracted from the palette for better effective demonstration and is located in the bottom section of this figure.

Phase-field models tend to be highly nonlinear so that its output can differ qualitatively, depending on the region in the input/parameter space where the sample is taken. Phase-field models are also highly complex in its formulation and are thus not amenable to intrusive approaches of UP. Moreover, they are computationally expensive, with full three-dimensional realizations of the simulations requiring upwards of 10,000 CPU-hours in some of the fastest supercomputers available.  Finally, the input/parameter space is high-dimensional, with more than $>$20 stochastic input conditions (i.e. temperature-time) and model parameters that dramatically affect the thermodynamic and kinetic state of the system, as shown in Figure~\ref{fig:mic_schematic}. 
    
\begin{figure}[!ht]
        \centering
        \subfloat[]{\includegraphics[width=0.3\columnwidth]{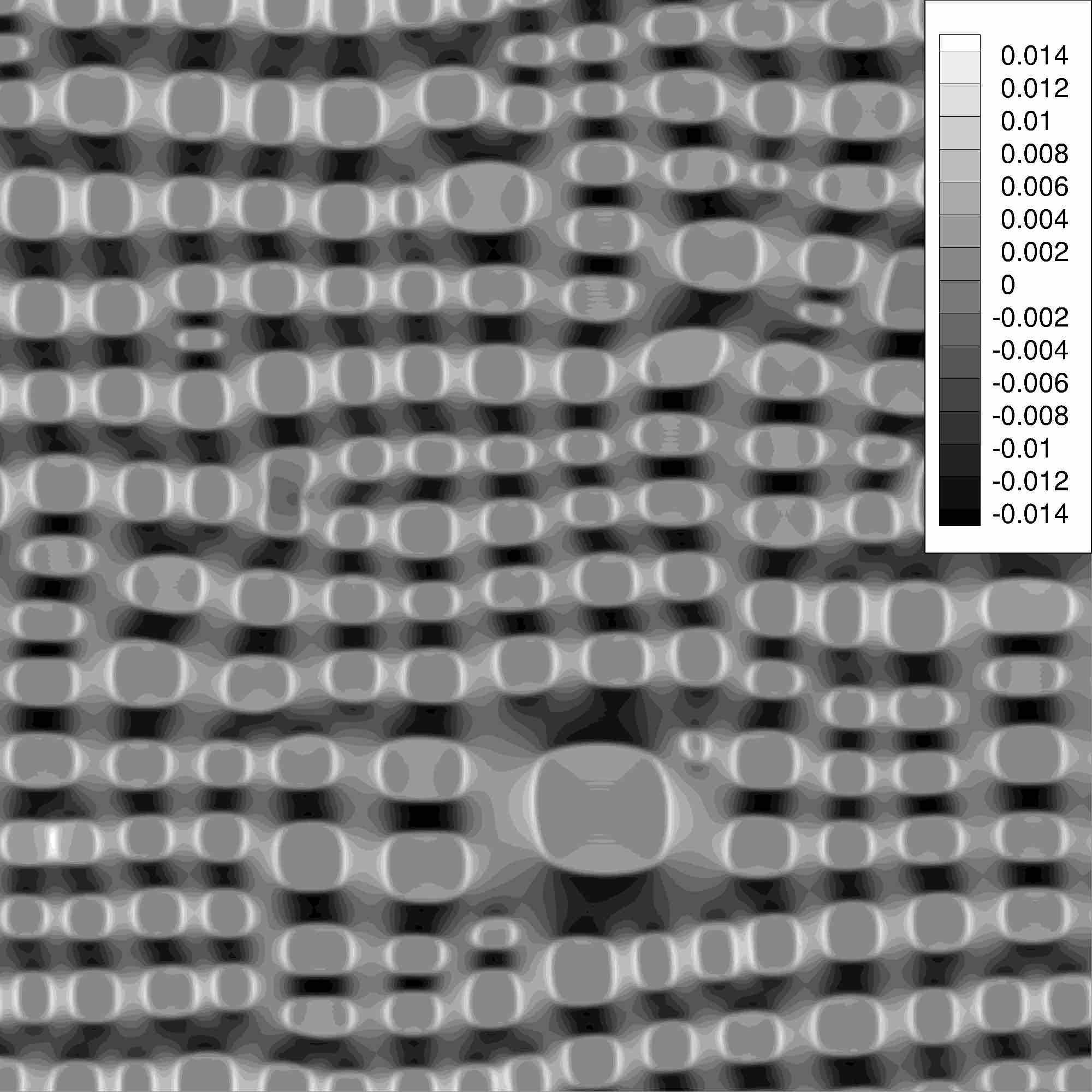}}  \hspace{0.15mm} 
        \subfloat[]{\includegraphics[width=0.3\columnwidth]{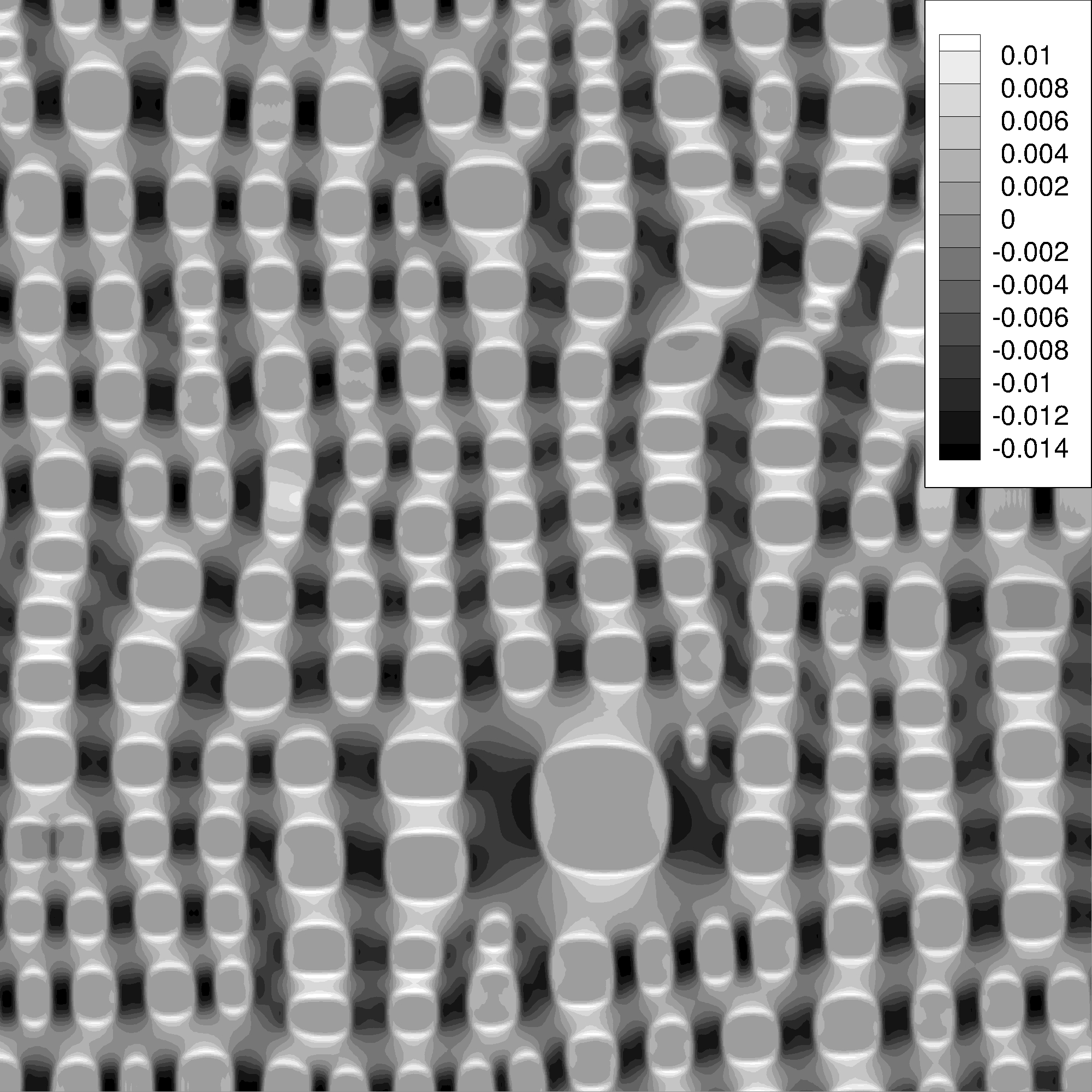}}  \hspace{0.15mm}
        \subfloat[]{\includegraphics[width=0.3\columnwidth]{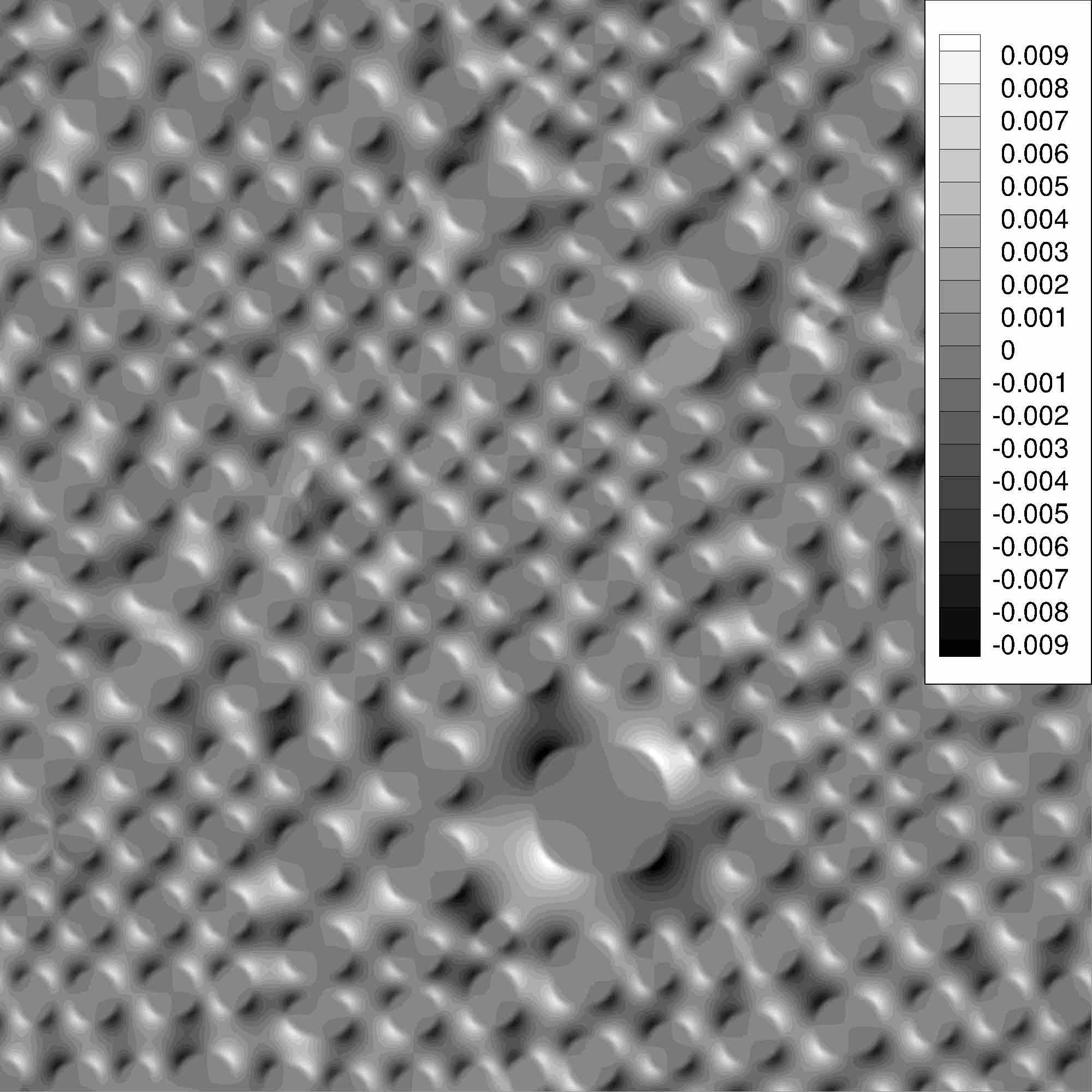}}\\\vspace{-0.35cm}  
        \subfloat[]{\includegraphics[width=0.3\columnwidth]{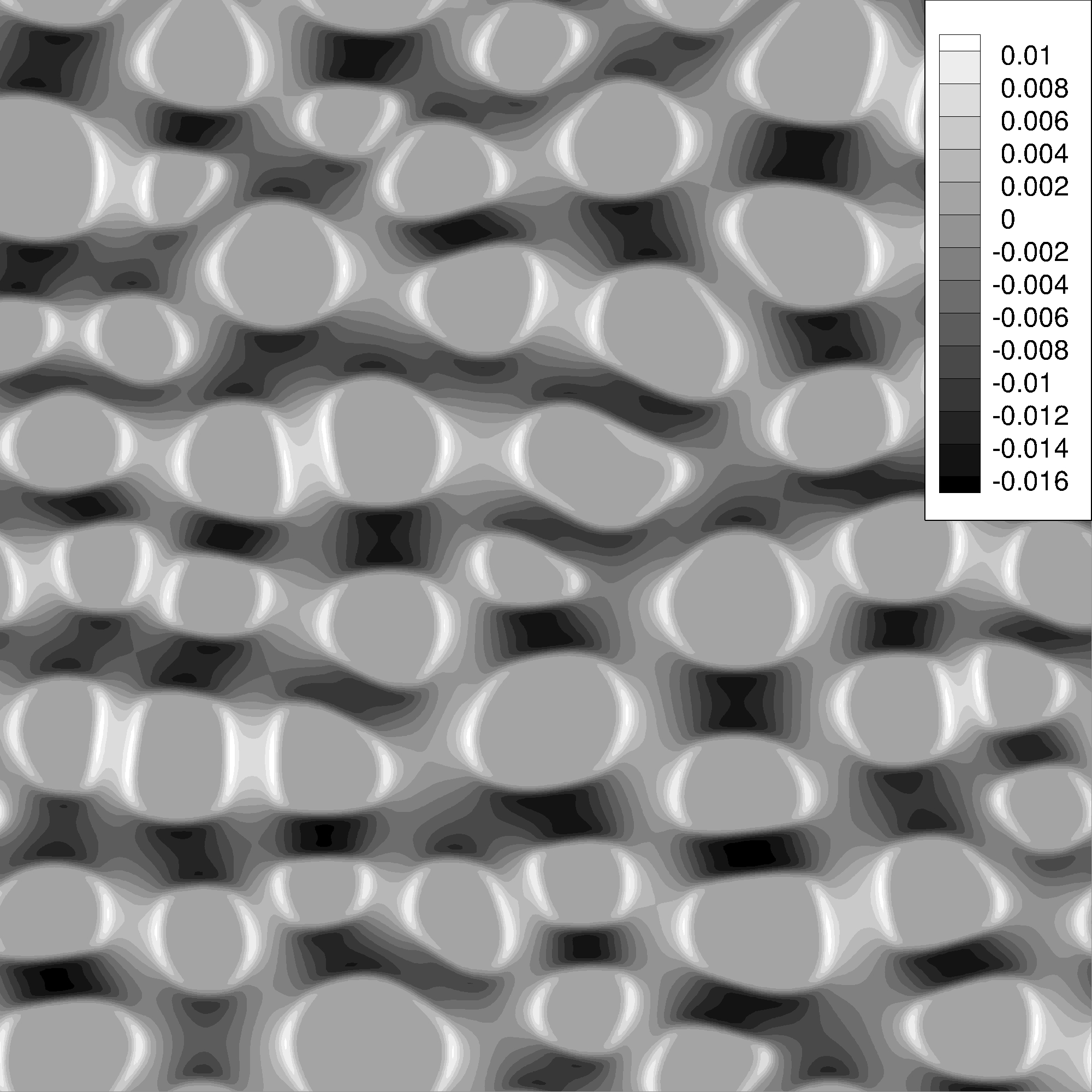}} \hspace{0.15mm} 
        \subfloat[]{\includegraphics[width=0.3\columnwidth]{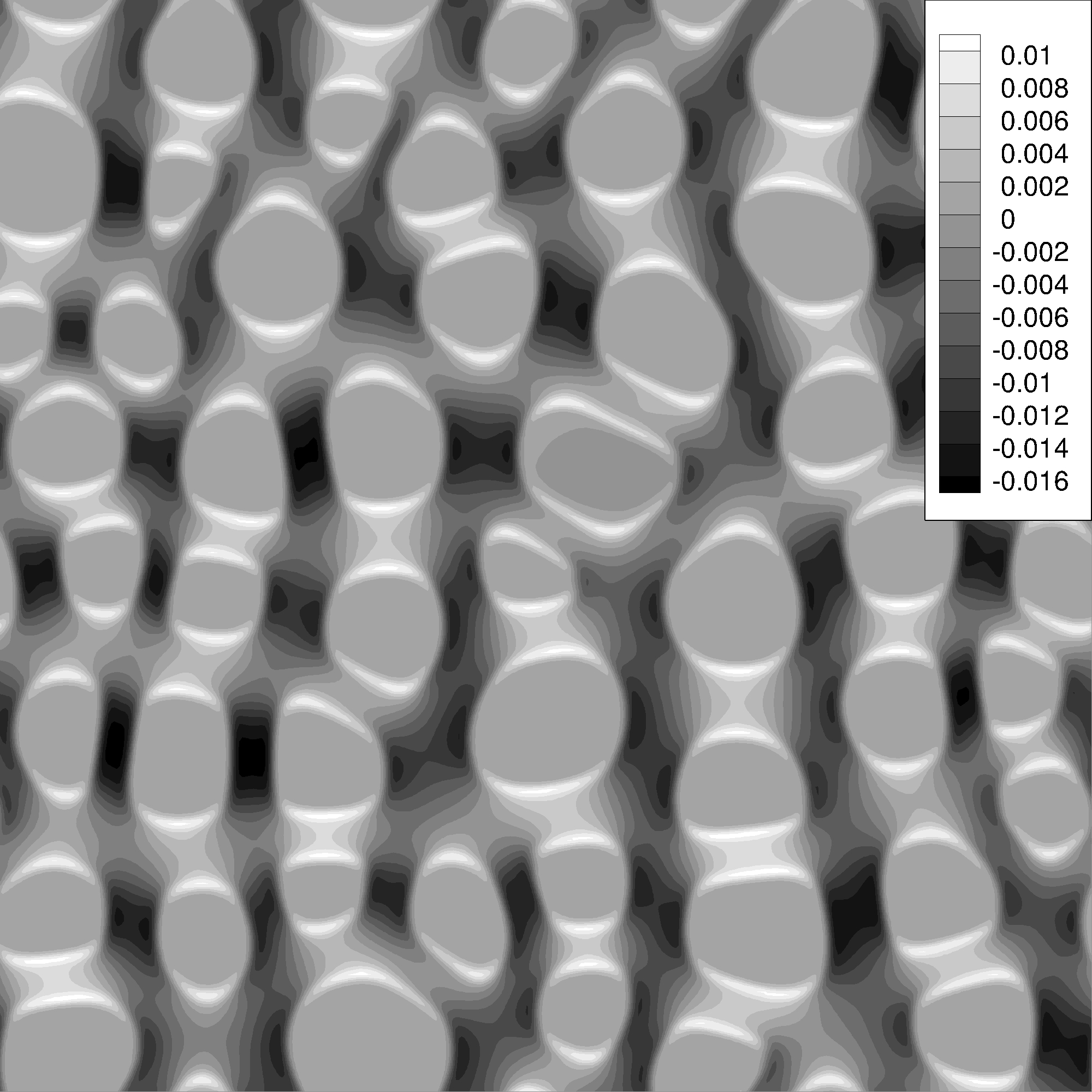}} \hspace{0.15mm}  
        \subfloat[]{\includegraphics[width=0.3\columnwidth]{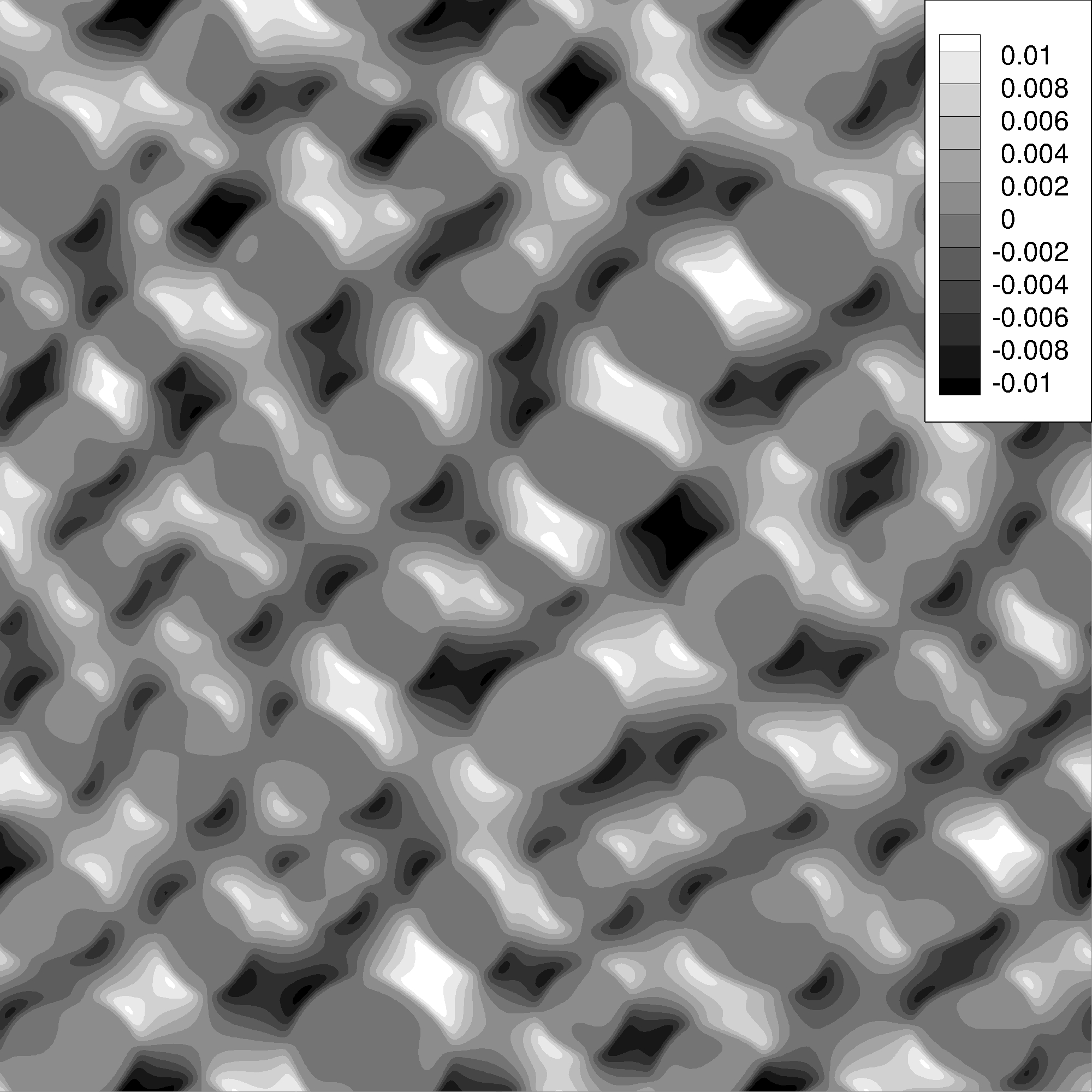}}         
        \caption{Effect of the local strain due to inhomogenous elastic effects in the microstructure. a and d) $\varepsilon^{el}_{11}$, b and e) $\varepsilon^{el}_{22}$, c and f) $\varepsilon^{el}_{12}$. All input parameters are different in these two cases. The SFTS value is $\varepsilon_{ii}^0=-0.0176$, elastic constants are $\left[Mg_2Si: C_{11}=120.2, C_{12}=22.1, C_{44}=56.3 \text{GPa} \right]$, $\left[Mg_2Sn: C_{11}=76.9, C_{12}=19.2, C_{44}=38.4 \text{GPa} \right]$ for the case in the first row, and $\varepsilon_{ii}^0=-0.0191$, $\left[Mg_2Si: C_{11}=120.5, C_{12}=22.2, C_{44}=48.8  \text{GPa} \right]$, $\left[Mg_2Sn: C_{11}=79.14, C_{12}=38.2, C_{44}=21.5 \text{GPa} \right]$ for the case in the second row.}
        \label{fig:strain_stress_map}
        \vspace{-0.2cm}
    \end{figure}

    
It must be noted that since we are interested in the elastochemical interactions in the microstructure, the effect of variations in elastic constants on the microstructure were considered. The local long-range interactions (i.e. strain and/or stress fields) of the heterogeneous multi-phase nanostructure are quite sensitive to the selection of the elastic parameters, and must be considered with a great care. Figure~\ref{fig:strain_stress_map} demonstrates the elastic strain maps ($\varepsilon_{ij}^{el}$) for two distinct microstructures with different set of elastic constants (i.e. shear constants), and very similar eigenstrain ($\varepsilon^0_{ii}$) values. Both cases can be categorized as cuboid-type microstructures. Though in the first case, (Fig.~\ref{fig:strain_stress_map}a,b, and c) the morphology is perfectly cuboid with a deviation from unimodal particle-size distribution. In the second case (Fig.~\ref{fig:strain_stress_map}d,e, and f), the particles are sheared at the corners to evolve towards a cuboidal shape during the coarsening stages as well. While microelastic effects can be considered to be very similar, slight differences in other parameters of the chain model yielded different observations.

\subsection{Microstructure Quantification}

Uncertainty propagation consists of linking the uncertainty in the input conditions and model parameters to the generated uncertainty in the model output. To properly quantify the uncertainty and study the way it propagates across the simulations, it is necessary to define quantity(ies) of interest (QoI) whose distribution can then be interpreted as a measure of variance resulting from the stochastic nature of the input space.
    
A traditional way to quantify a microstructure is to exploit several QoI in the (micro)structure. A challenging aspect of the uncertainty propagation effort in this work is the large dimensionality of both the input and the output space. The large dimensionality of the input space originates from the many parameters that are necessary to complete the phase field model. Given the fact that the microstructure space is highly heterogeneous, it is to be expected that many QoIs would be necessary to complete characterize the effect of model inputs on the morphology and topology of the resulting microstructures. In order to handle these very complex spaces, we will resort to machine learning frameworks, as will be described below.

The propagation of uncertainty first requires the quantification of changes in the microstructure. Given the large dimensionality of the output (512x512 pixels), dimensional reduction is necessary. Moreover, in order to enhance the interpretability of the analysis it is desirable for the reduced dimensions to have physical significance. The determination of QoIs typically involves extraction of features or disclosing a bank of descriptors that can be used to train a classifier based on the frequency of observations \cite{karsanina2015universal,fullwood2008microstructure}. A conventional yet very useful approach is to use semantic texton forests \cite{shotton2008semantic}. This is specifically useful when the phase-field variable is composition and can also be broadly used in the real-image data. In addition, visual words \cite{csurka2004visual} can be used as powerful discrete image representations for categorization. Another way to tackle this is to utilize one or a combination of Filter-bank responses (e.g. Fourier or other sort of wavelets), and one or combination of invariant descriptors (e.g. SIFT \cite{lowe2004distinctive,ham2019automated}). 

\begin{figure}[h!]
        \centering
        \subfloat[]{\includegraphics[width=0.28\columnwidth]{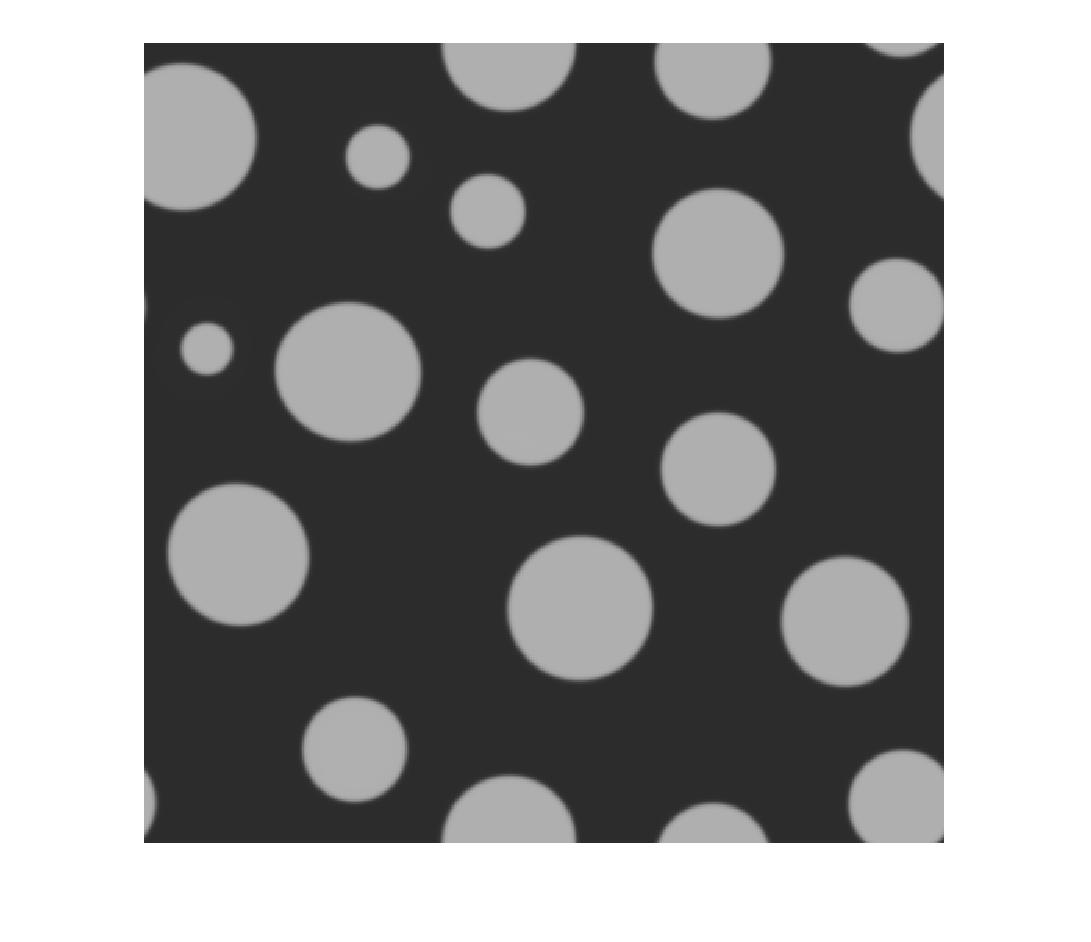}} \hspace{-0.45cm}
        \subfloat[]{\includegraphics[width=0.28\columnwidth]{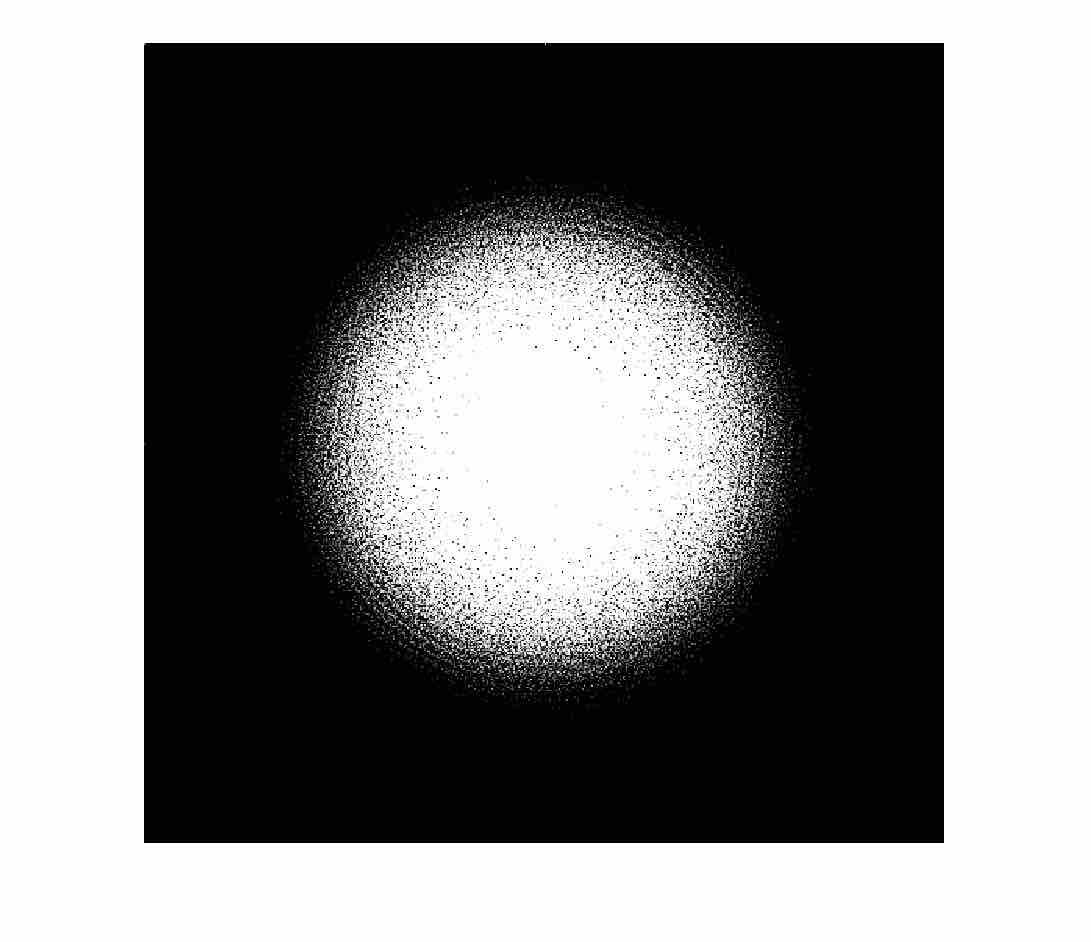}} \hspace{-0.45cm}
        \subfloat[]{\includegraphics[width=0.28\columnwidth]{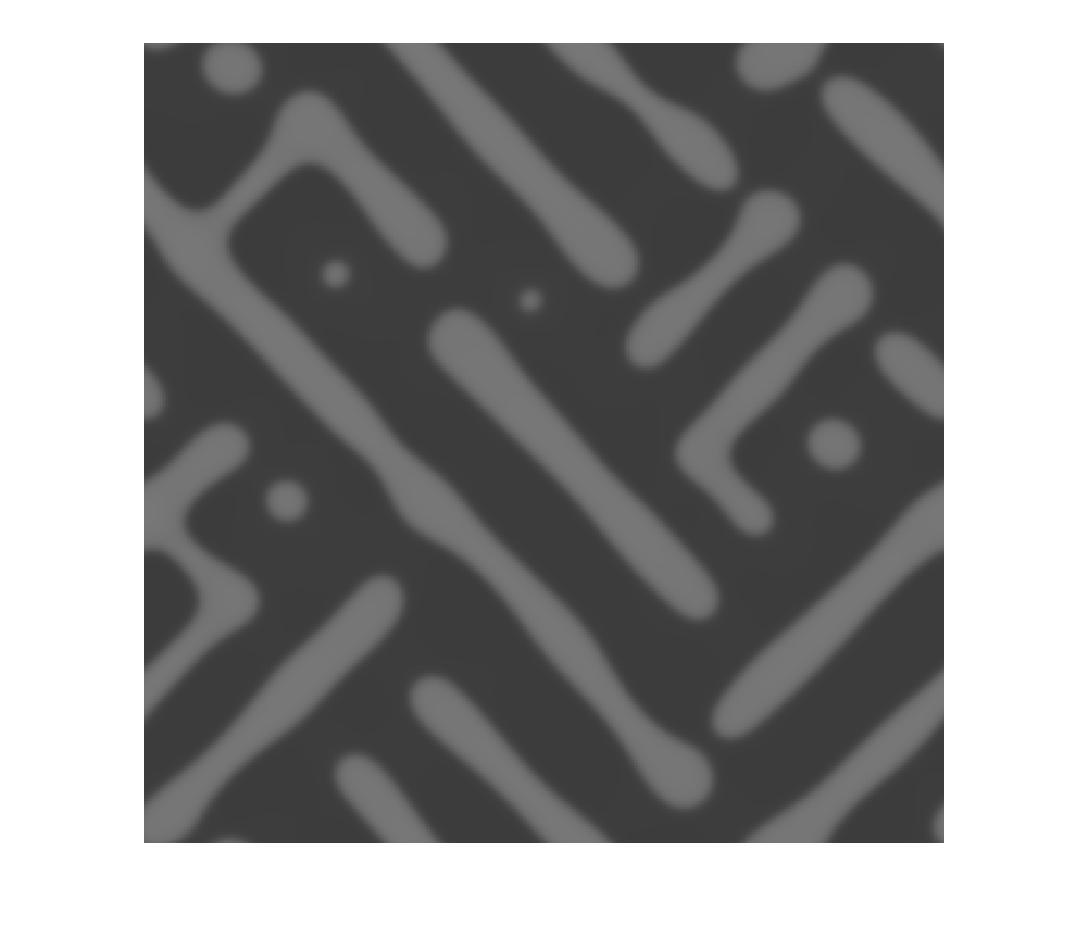}} \hspace{-0.45cm}
        \subfloat[]{\includegraphics[width=0.28\columnwidth]{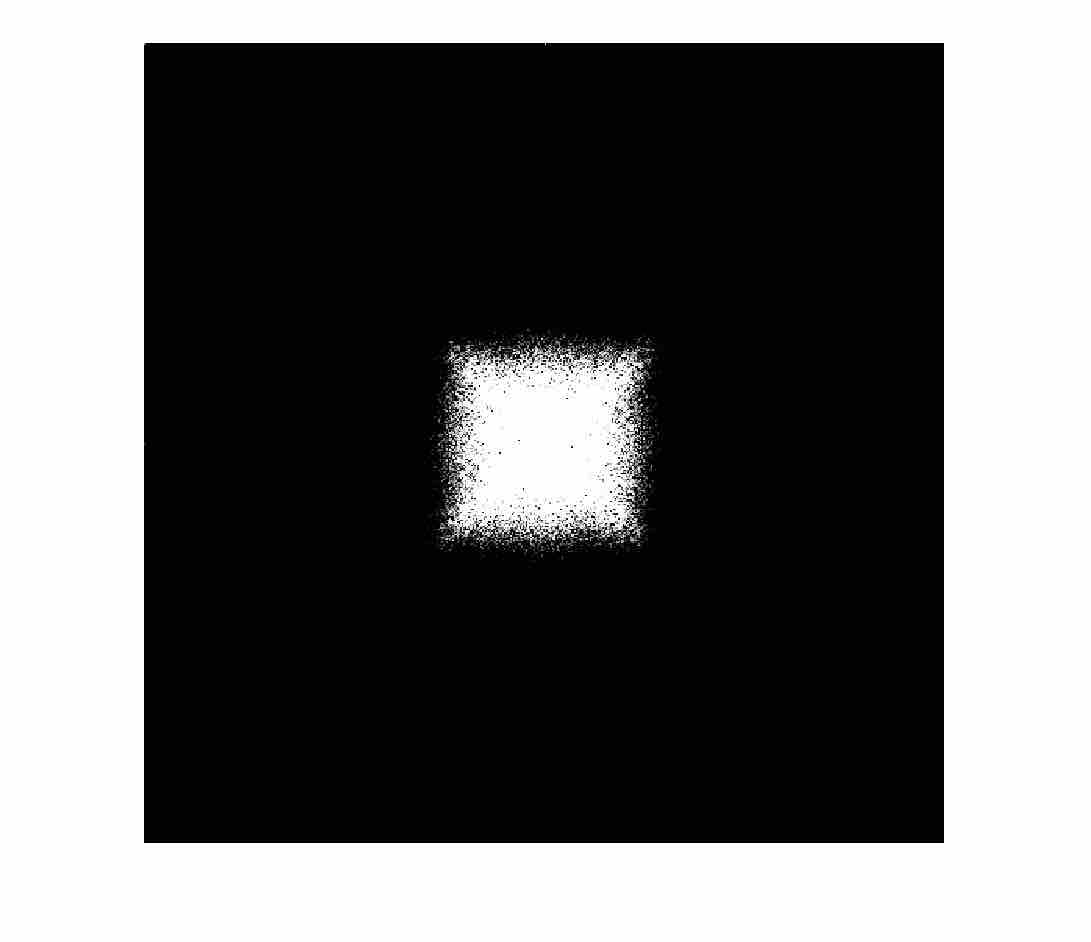}} \hspace{-0.45cm} \vspace{-0.2cm}  
        \caption{a,c) Example microstructures, and the corresponding b,d) Power spectrum maps obtained by $Fourier$ transformation of the microstructures generated from the elastochemical phase field model.}
        \label{fig:mic_fourier}
    \end{figure}

A series of microstructure analytic tools were used to determine QoIs to evaluate the developed UP framework. We use metrics such as average feature size, area fraction, composition of the phases, aspect ratio as well as increasingly common approaches such as n-point statistics~\cite{fullwood2008microstructure}, frequency-domain analysis (c.f. Figure~\ref{fig:mic_fourier}). In the later case, a general information about the morphology and orientation of the particles are reflected in the frequency space. Many of the commonly used metrics provide information about the average state of microstructure spaces, but in many cases they do not provide information about their topology. In the case of transport behavior (such as phonon conductivity), topology may play an important role and metrics quantifying this microstructure feature are necessary.

    
Table \ref{tab:posterior_dist} summarizes the list of QoIs and the associated statistics for phase-field model outputs. It contains the posterior, means, index of dispersion denoted by variance to mean ratio (VMR) of the obtained QoIs. VMR is an index to quantify the dispersion of a probability and measures the clusterability and variability in the data. For VMR$>1$, and $0<$VMR$<1$ the data are over-dispersed and under-dispersed, respectively. VMR = 0 is associated to a random data-set and VMR = 1 corresponds to a Poisson distribution. Except the last three, all other QoIs demonstrate multimodal distributions. We wish to note that this multi-modality would render simplified uncertainty analysis frameworks---such as those based on sampling of min, max and mean values of input parameters---ineffective. 

It should be noted that C$^{\alpha}$ and and C$^{\beta}$ are considered to be the equilibrum compositions of Mg$_2$Sn and Mg$_2$Si phases, respectively. Both of these demonstrate a trimodal distribution: I) the first peak belong to the case where the process of phase separation has entirely finished and the elastic interactions are weak. In this case, we obtain the equilibrium values corresponding to the sampled phase-diagram. II) In the second case, the composition between elastic and chemical driving forces are great, and morphology of the particles are similar to those reported in elastochemical interactions. III) In the third case, either kinetics is slow or the system tends to dissolve rather than phase separate. In such cases, the elastic driving forces are significantly stronger than the chemical ones. This is believed to be achieved by the non-equilibrium synthesis conditions (refer to \cite{yi2018strain} for further information). 

    \begin{table}[h!]                                       
        \centering 
        \scriptsize
        \caption{List of extracted QoIs and their posterior distributions, mean, standard deviation, minimum and maximum from the entire phase-field runs.}                            
        \begin{tabular}{m{1.75cm}m{3.4cm}m{1.1cm}m{1.6cm}@{}m{0cm}@{}}  
        \toprule                                                
        Target variables & Posterior Distribution & $\mu$ & Dispersion ($\frac{\sigma^2}{\mu}$)  \\    
        \midrule 
        $C^{\alpha}$    & \includegraphics[width=0.29\columnwidth]{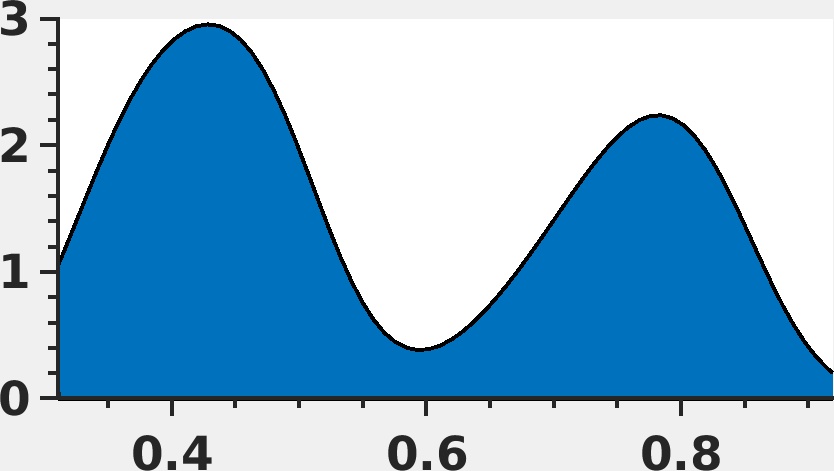}   & 0.57 & 0.056   \\     
        $C^{\beta}$     & \includegraphics[width=0.29\columnwidth]{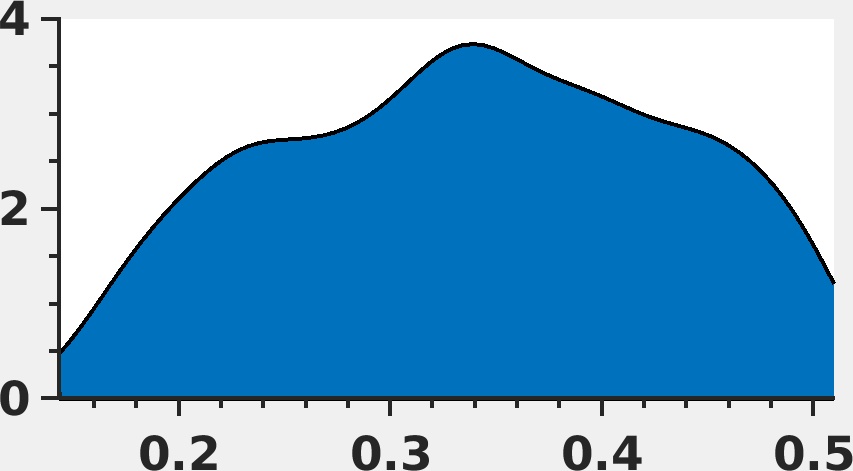}   & 0.34 & 0.026   \\     
        Char. Length    & \includegraphics[width=0.29\columnwidth]{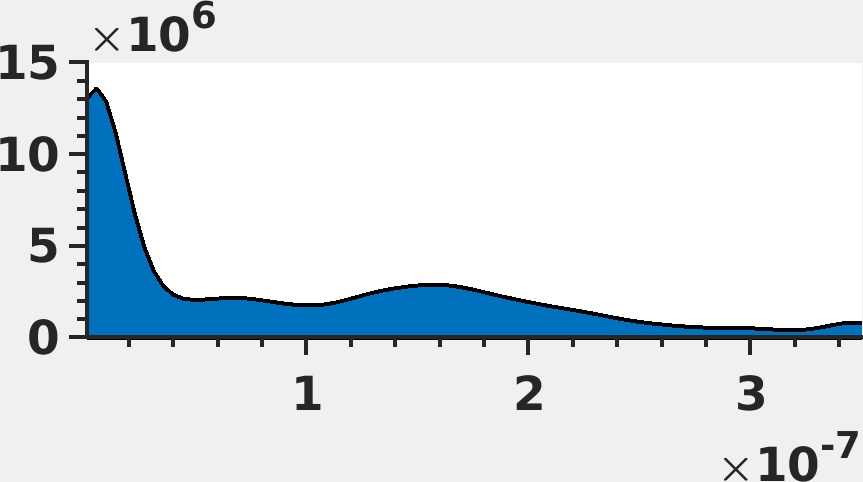}   & 8.62$\times$10$^{-8}$ & 1.08$\times$10$^{-7}$     \\    
        Area fraction   & \includegraphics[width=0.29\columnwidth]{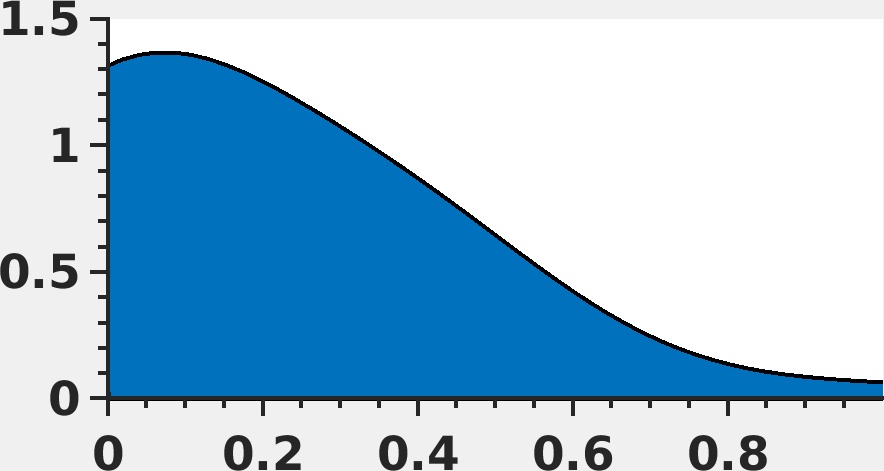}   & 0.17 & 0.30 \\    
        Roundness       & \includegraphics[width=0.29\columnwidth]{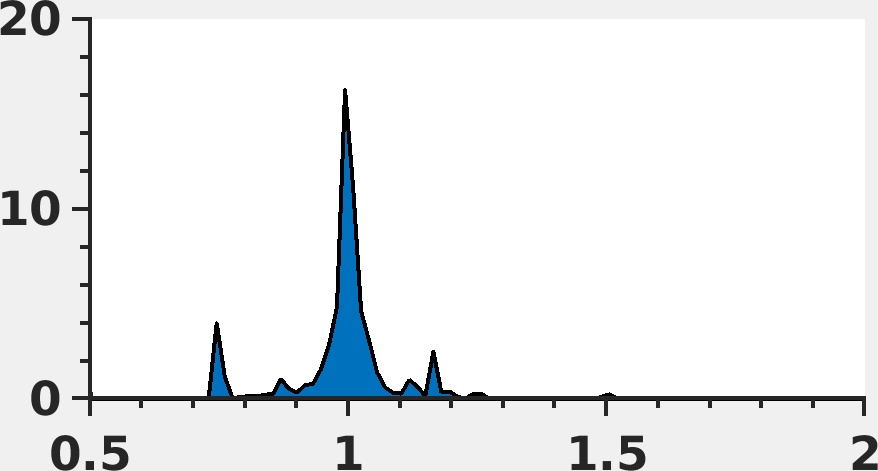}   & 0.99 & 0.70   \\    
        \normalsize{$\sfrac{\text{Diagonal}}{\text{width}}$}  & \includegraphics[width=0.29\columnwidth]{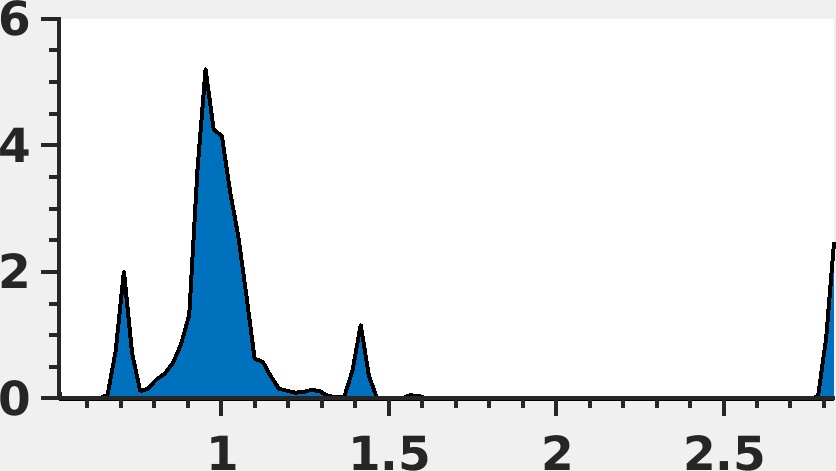}   & 1.18 & 0.29   \\      
        \normalsize{$\sfrac{\text{Diagonal}}{\text{height}}$} & \includegraphics[width=0.29\columnwidth]{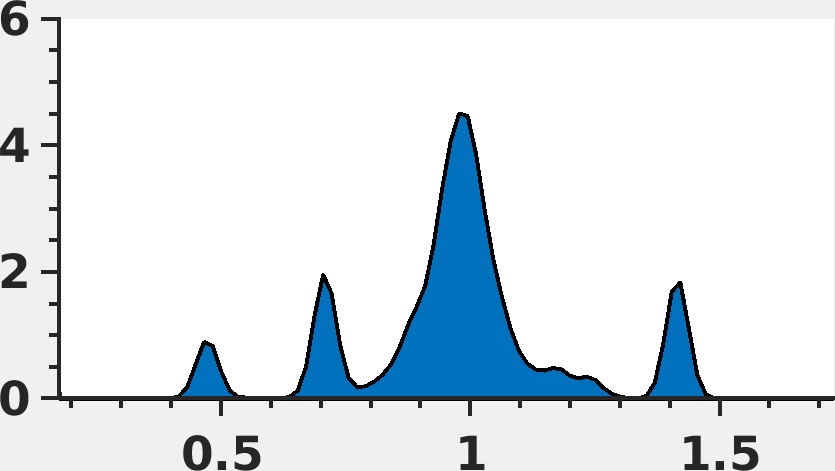}   & 0.98 & 0.05  \\     
        $\bar{\mu}_{chem}$  & \includegraphics[width=0.29\columnwidth]{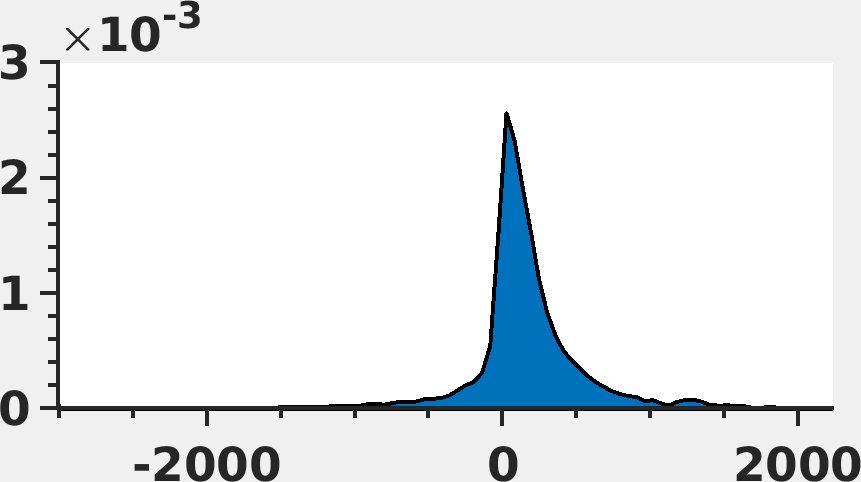}   & -156.06 & -1012.1  \\     
        $\bar{\mu}_{elas}$  & \includegraphics[width=0.29\columnwidth]{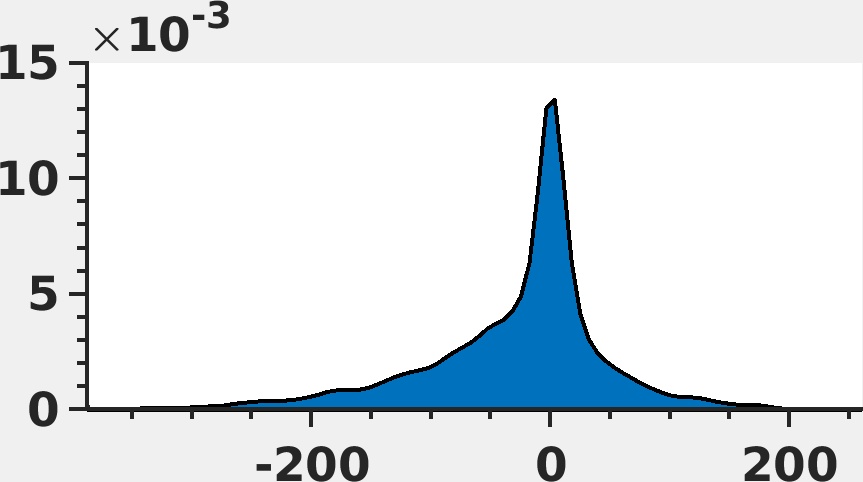}   & -27.83 & -214.75  \\     
        $\bar{\mu}_{int}$   & \includegraphics[width=0.29\columnwidth]{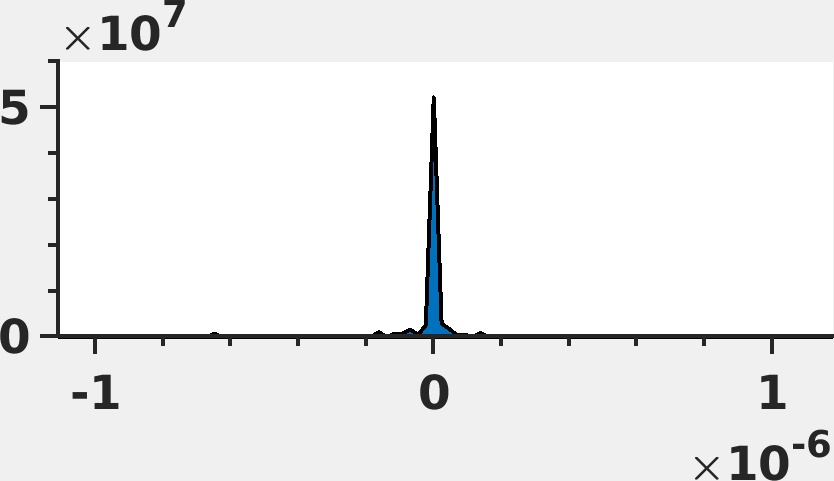}   & 1.59$\times$10$^{-9}$ & -6.01$\times$10$^{-6}$   \\    \bottomrule 
        
    \end{tabular}                                       
    \label{tab:posterior_dist}                          
    \end{table}     
    
    
The characteristic length scale for each microstructure is obtained by calculating the radial basis power spectrum of the Fourier transform of the microstructure (2D composition space). The distribution is bimodal and its peaks belong to either the not-decomposed, or the decomposed structures with certain morphologies. The calculated area fractions uniformly ascends from zero fraction for dissolved or not-decomposed microstructures to the decomposed microstructures. The probability density of observation of not-decomposed or dissolved microstructures are higher than its counterpart for the decomposed ones. Roundness is extracted based on the shape of the morphology that appears in the power spectrum domain. Figure~\ref{fig:mic_fourier} demonstrates examples of different power spectrums that are obtained for the shown microstructures. For circular morphology the power spectrum is symmetric and appears to be a circle with a diameter based on the coarsening stage. The power spectrum appears to be in the form of rectangle/square for a perfectly oriented morphology. 
    
Quantities $\sfrac{Diagonal}{width}$ and $\sfrac{Diagonal}{height}$ are the relative ratio of the diagonals of the power spectrum centroid with respect to its width and height. The closer the value is to $\sqrt{2.height^2}$ and/or $\sqrt{2.width^2}$, the morphologies tend to shear more. Moreover, the higher these values get, the anisotropy becomes more evident. On the other hand, the mean value of each of the contributing energy fields are also obtained for the current sampling. The mean of bulk and elastic driving forces in the domain demonstrate a unimodal probability distributions. However, the mean of interfacial driving forces demonstrates multimodal isolated peaks. 

Table \ref{tab:posterior_dist} represents the \textbf{total variance} in the microstructure space assuming the CDF of the input parameters and their statistical correlation. The propagated uncertainty is represented through different QoIs. However, these results are aggregated and further analysis is necessary in order to examine the effect of individual parameters (or sets of parameters) on the resulting microstructures. The large dimensionality of the input and output spaces, as we discussed above, makes it necessary to rely on machine learning approaches that facilitate the analysis of the microstructure space and their relation to model inputs.

    \subsection{Application of the Materials Informatic Techniques in Microstructural Evaluation}
    
    \textcolor{black}{
    A primary purpose of data-mining techniques is to help in determination of the possible patterns for better prediction abilities, which form the foundations for understanding materials behavior~\cite{rajan2013informatics}. Materials informatics is an interdisciplinary blend of statistics, machine learning, artificial intelligence, pattern recognition and materials science. Here we employ a few core tasks (e.g. Cluster analysis, Anomaly detection, etc.) on the posterior data generated in this study. This collective integration of statistical learning tools with experimental and/or computational materials science allows for an informatics driven strategy for materials design \textcolor{black}{and development under the framework of ICME.}}
    
    \begin{figure}[!ht]
        \centering
        \includegraphics[width=0.95\columnwidth]{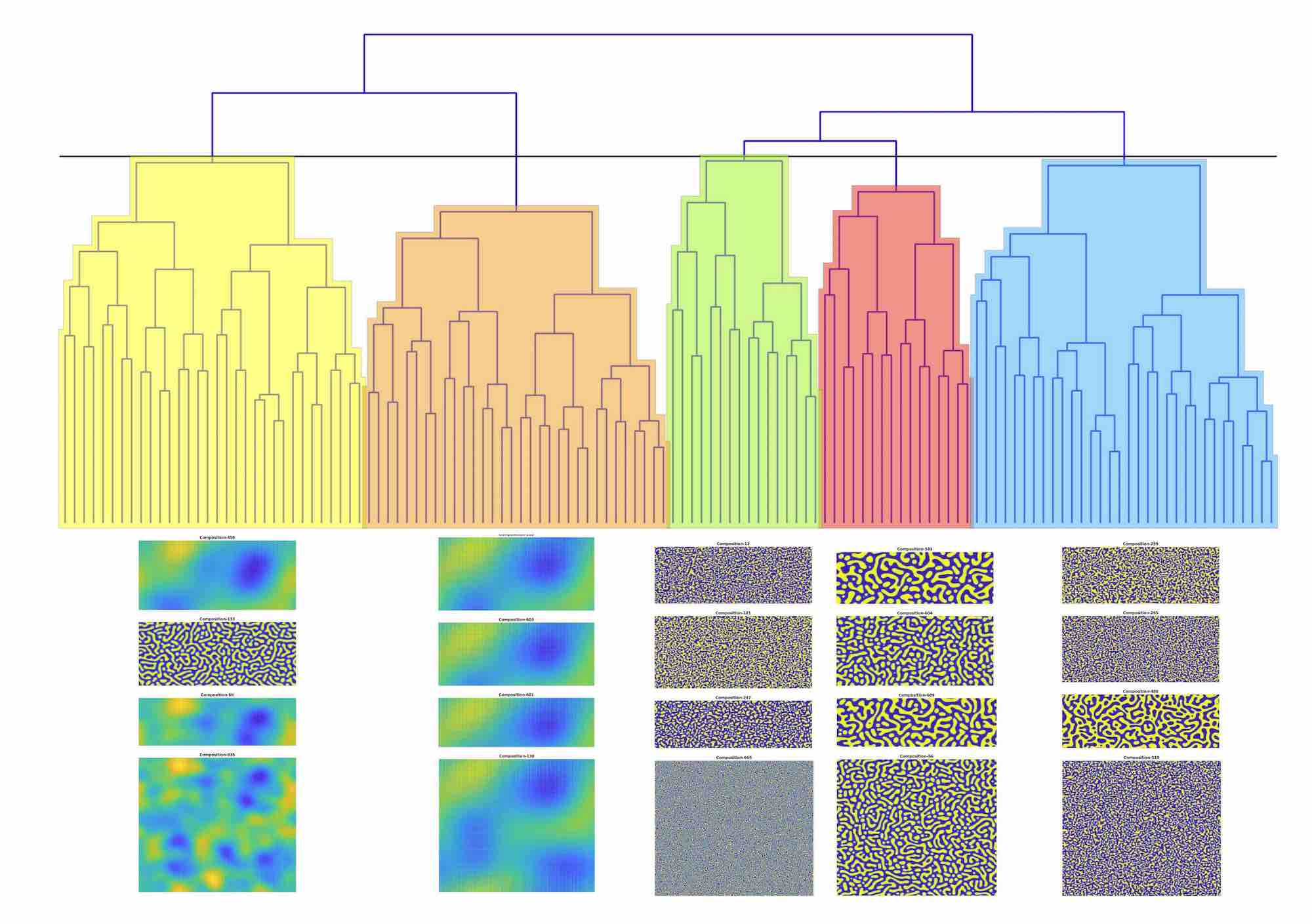}\\
        \caption{Hierarchical clustering using Mahalanobis distance matrix. The horizontal line indicates 78\% height ratio. The images are clustered into two main subcategory of `decomposed' versus `not-decomposed' microstructures. Each of these main categories are also divided into three sub-levels with an acceptable error.}
        \label{fig:dendrogram}
    \end{figure} 
      
A preliminary simple hierarchical clustering using different distance metrics are performed to elucidate the structure of the available microstructural data. The results are demonstrated as a corresponding dendrogram in Fig.~\ref{fig:dendrogram}. This calculation is based on the the 18 material parameter as the inputs and the eight QoI as the target variables. In addition, several distance metrics (i.e. Euclidean, Manhattan, Mahalanobis, Spearman and Pearson) are used. The result in Fig.~\ref{fig:dendrogram} is based on Mahalanobis distance metric where the elements are taken in a pairwise fashion between the elements of a given set using $d_M(x,y) = \sqrt{ (x - y)^T S^{-1} (x - y) }$ where $S$ is the Covariance matrix. The dendrogram shown in Fig.~\ref{fig:dendrogram} is created by the Ward's linkage criterion (increase in variance for the cluster being merged) and aims to indicate the similarity/dissimilarity among annotation categories. The five sub-clusters of the dendrogram shown in different colors are annotated with selected representing microstructures. As seen, this dendrogram clusters the microstructures into two main categories, i.e. `decomposed' and `not-decomposed' classes, with an acceptable error. 

    
A comparison of several classifiers is performed on the extracted QoIs data-set. Classification methods are suited for cases where the class label is discrete. Hence, we simply use the `not-decomposed' versus `decomposed' class labels. Here we return to the connection with performance in TE materials by noting that microstructures that correspond to non decomposed states could be associated with alloying/mass phonon scattering, while decomposed microstructures corresponds to interfacial phonon scattering. The length scales of different scattering mechanisms are different and are thus expected to change the phonon transport characteristics and the corresponding thermoelectric performance of the Mg$_2$\{Sn,Si\} system~\cite{tazebay2016thermal}. 

Figure~\ref{fig:classification} illustrates the nature of such decision boundaries in the 2D input parameter regions which can be used to determine the desired regions (alloy/mass vs interface scattering) in the material parameter space. Several classifiers are tested on all of the parameter pairs and the selected classifiers to report are the Nearest neighbors, Gaussian process, Radial Basis Function (RBF) kernel Support Vector Machine (SVM), Neural Network. Each subfigure in Fig.~\ref{fig:classification} illustrates the nature of decision boundaries for a given pair of the input parameter using different classifiers. The training points in these plots are shown in solid colors and the testing points are defined by the semi-transparent contour. We first randomly extracted 10\% of the original data obtained from a set of microstructures frozen at a fixed time. 60\% of this data is used as the training set and 40\% as the test set. Figure~\ref{fig:classification}a demonstrates the Nearest neighbors classifier in the (alloy composition ($X_{Si}$), $a_{SS}^0$) space with a transitioning boundary where the points are mixed in the boundary of the two class. Figure~\ref{fig:classification}b demonstrates the Gaussian process classifier  \cite{williams2006gaussian} result in the ($\varepsilon^T$, $a_{SS}^0$) space with a smooth, and continuous separation boundary between the decomposed (blue) and not-decomposed (red) clusters. Figure~\ref{fig:classification}c shows the classification result in (C$_{11}$:Mg$_2$Si, $a^0_{SS}$) space using RBF SVM classifier that reveals the transitioning bounderies between the two clusters. The red cluster preserves a larger area while certain points of this cluster are mixed with the blue cluster. \textcolor{black}{The result associated with the trained neural network classifier in ($\kappa$, $a_{SS}^0$) space is shown in Fig.~\ref{fig:classification}d, which again illustrates a smooth, and continuous separation boundary between the two classes. However, it should be noted that no clear classification boundaries are obtained in most pair-parameter spaces which can be attributed to the lack of recognition through 2D projections of the high dimensional parameter space. We note that in these cases, the application of a dimensional reduction technique, e.g. principal component analysis (PCA), might be helpful for more rigorous classification study.}



    \definecolor{red}{HTML}{FF0000}
    \definecolor{blue}{HTML}{0000FF}
\begin{figure*}[ht!]
    \centering
    \subfloat[]{\includegraphics[width=0.24\textwidth]{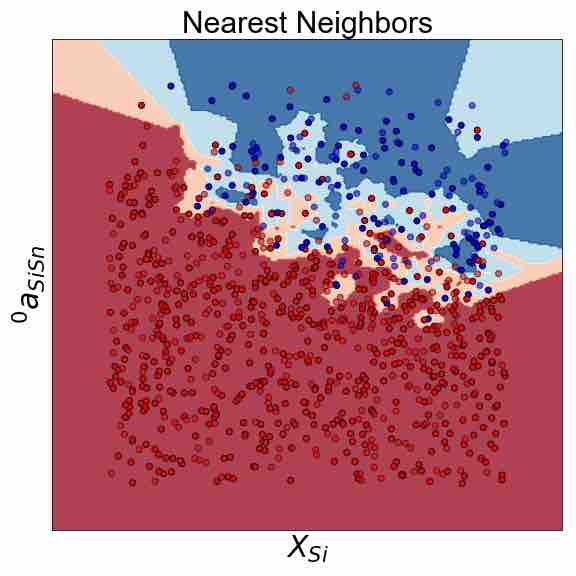}}
    \subfloat[]{\includegraphics[width=0.24\textwidth]{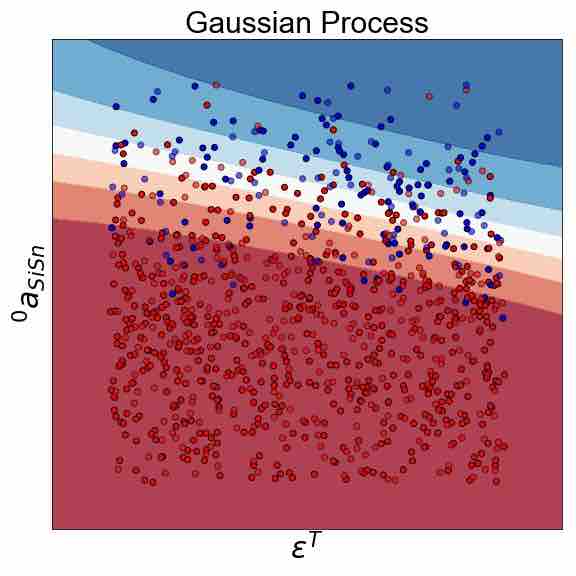}}
    \subfloat[]{\includegraphics[width=0.24\textwidth]{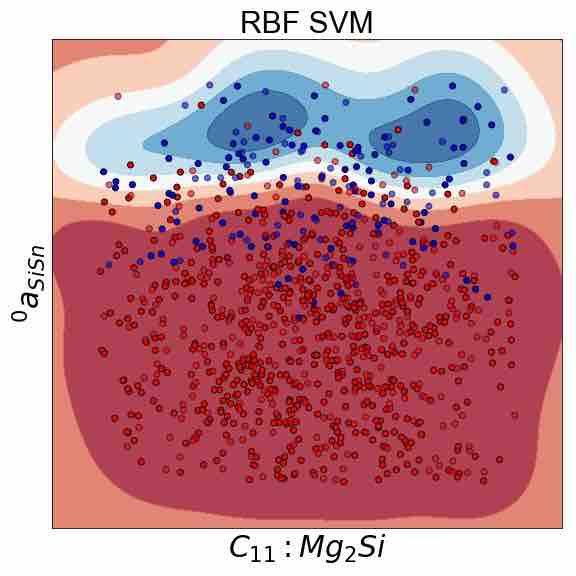}}
    \subfloat[]{\includegraphics[width=0.24\textwidth]{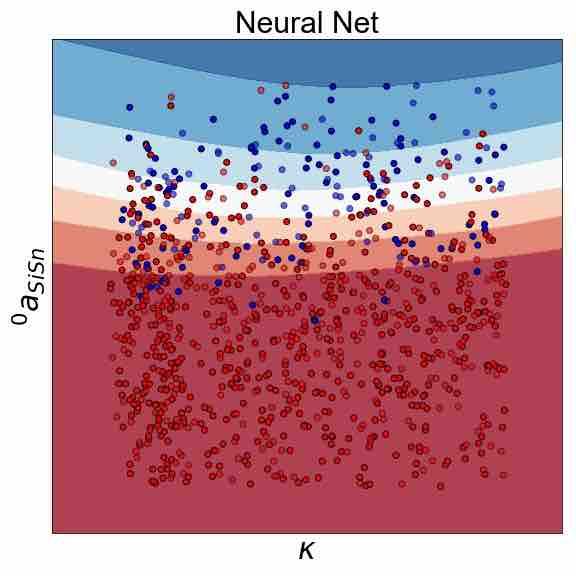}}

    \caption{Classification results for a) Nearest neighbors classifier in (alloy composition ($X_{Si}$), $^0a_{SS}$) space, b) Gaussian process classifier in ($\varepsilon^{T}$, $^0a_{SS}$) space, c) Radial basis function support vector machine (RBF SVM) classifier in ($C_{11}$:Mg$_2$Si, $^0a_{SS}$) space d) Neural network classifier in ($\kappa$, $^0a_{SS}$) space. For interpretation of the colors the reader is referred to the online version. \raisebox{1mm}{\fcolorbox{black!90!gray}{red}{\null}} stands for not-decomposed or dissolved microstructures, while \raisebox{1mm}{\fcolorbox{black!90!gray}{blue}{\null}} stands for decomposed microstructures with different morphologies.}
    \label{fig:classification}
\end{figure*}

While further investigation of the microstructure space through machine learning approaches may be warranted in order to fully characterize the connections between input parameters and resulting microstructures, this is well beyond the scope of the present work. It is in fact our intent to further explore the use of different machine learning frameworks into the obtained microstructure dataset.

    \section{Summary and Conclusion}\label{sec:conclusions}
    
As with any computational analysis, uncertainty quantification/propagation plays a major role in predicting the outcome of multi-scale models in materials science. One of the most important tasks in the materials design under ICME framework is propagation of uncertainties of parameters across the multi-scale models that connect process-structure-property-performance. Hence, we addressed an existing challenge in materials science, i.e., UP from the thermodynamic parameters to the microstructural features through a chain of CALPHAD, microelasticity, and phase field models. This is a multi-step uncertainty propagation with the techniques used in each step motivated by the type of problem that is being dealt with. Technically, thermodynamic parameters in the CALPHAD model and their underlying imposed uncertainties obtained from an MCMC sampling approach has been propagated to Gibbs free energy of phases and equilibrium phase diagram through a forward analysis of an ensemble of these samples. Then, these uncertainties as well as the uncertainties of microelastic and kinetic parameters have been propagated to the microstructural features using a Gaussian copula sampling approach.  
    
While the present study has multiple aims, one of these is to rectify the common belief towards the deterministic assumption about the parameter values in the phase field models as well as the experimental data used for the calibration process.


    
The propagation of uncertainty in the prior parameter space using model chains resulted in a massive microstructure data-set ($\sim$50TB). The quantities of interest in the microstructures are identified and extracted using automated frameworks from the large amounts of data that is generated by HT-phase-field runs. We used eight QoIs to map the obtained probability distributions of parameters into probability distributions of the extracted quantities. Then, data-mining techniques are employed to find patterns in the parameter space that can contribute to better understanding of process-microstructure relations. \textcolor{black}{The results show that the data is clustered into two main categories of `decomposed' and `not-decomposed' microstructures. These results can be very useful for engineering material behaviour in favor of specific phonon scattering mechanism  and/or better thermoelectric response. The proposed framework is generalizable to applications to other materials problems and microstructure-sensitive properties. As of this writing further exploration of the input parameter-microstructure space is in progress and will be reported in our future work.}

    
Developing general QoIs of the microstructural space will enable better quantification of the uncertainty propagated through the models establishing inverse maps to connect regions in the microstructure space to the corresponding regions in the input space. While standard microstructure analysis approaches can be used to represent microstructure spaces in a compact way. To establish PSP connections, it is very important to predict properties that depend on the specific features of such microstructures.

    \section*{Acknowledgment}
    The authors would like to acknowledge the \#Terra supercomputing facility of the Texas A\&M University, for providing computing resources useful in conducting the research reported in this paper. The data is currently hosted by this facility. This research was supported by the National Science Foundation under NSF Grant No. CMMI-1462255 and NSF-CMMI-1663130. RA and DA also acknowledge the support of ARL through [grant No. W911NF-132-0018].

    \section*{References}
    \bibliography{refs} 

\end{document}